\documentclass{article}

\usepackage{arxiv}

\usepackage[utf8]{inputenc} 
\usepackage[T1]{fontenc}    
\usepackage{hyperref}       
\usepackage{url}            
\usepackage{booktabs}       
\usepackage{amsfonts}       
\usepackage{nicefrac}       
\usepackage{microtype}      
\usepackage{doi}
\usepackage{subcaption}     
\usepackage{bbm}
\usepackage{multirow}

\usepackage{amssymb}
\usepackage{amsmath}
\usepackage{mathtools} 
\usepackage{xcolor}
\usepackage{media9} 
\usepackage{graphicx, animate}
\usepackage{units} 
\usepackage{esvect}
\usepackage{comment}
\usepackage{cite} 
\usepackage{float}
\usepackage{bm}

\usepackage[labelfont={bf,sf},font={small}, labelsep=space]{caption}
\usepackage{algorithm}
\usepackage{algpseudocode}

\usepackage{listings}

\definecolor{darkred}{rgb}{0.6, 0, 0}
\definecolor{softgreen}{rgb}{0.2, 0.6, 0.2}

\usepackage{todonotes}
\newcommand{\divergence}{\mathop\mathrm{div}}

\newcommand{\sY}{\hspace{-0.05cm} \prescript{S}{}{Y}}

\title{A Hybrid ABM-PDE Framework for Real-World Infectious Disease Simulations} 

\date{} 					

\author{ 
{
Kristina Kehrer}
\\
	Zuse Institute Berlin\\
	Berlin, 14195 \\
	\texttt{kehrer@zib.de} \\
	\And
 {
Tim O. F. Conrad} \\
	Zuse Institute Berlin\\
	Berlin, 14195 \\
	\texttt{conrad@zib.de} \\
}

\renewcommand{\headeright}{Paper}

\allowdisplaybreaks 

\begin{document}
\maketitle
\begin{abstract} 
This paper presents a hybrid modeling approach that couples an Agent-Based Model (ABM) with a partial differential equation (PDE) model in an epidemic setting to simulate the spatial spread of infectious diseases using a compartmental structure with seven health states. 
The goal is to reduce the computational complexity of a full-ABM by introducing a coupled ABM-PDE model that offers significantly faster simulations while maintaining comparable accuracy. Our results demonstrate that the hybrid model not only reduces the overall simulation runtime (defined as the number of runs required for stable results multiplied by the duration of a single run) but also achieves smaller errors across both 25\% and 100\% population samples. 
The coupling mechanism ensures consistency at the model interface: agents crossing from the ABM into the PDE domain are removed and represented as density contributions, while surplus density in the PDE domain is used to generate agents with plausible trajectories derived from mobile phone data.
We evaluate the hybrid model using real-world mobility and infection data for the Berlin-Brandenburg region in Germany, showing that it captures the core epidemiological dynamics while enabling efficient large-scale simulations. These results demonstrate that the proposed ABM-PDE framework provides a robust and computationally efficient alternative to full-scale agent-based simulations, making it suitable for realistic epidemic modeling and scenario analysis.
\end{abstract}


\section{Introduction}

Understanding and efficiently simulating complex dynamical systems is a central challenge across many scientific and engineering domains. In epidemiology, this challenge is particularly pronounced due to the need to capture diverse phenomena such as individual behavior, spatial mobility, and population heterogeneity. 
In this field, various modeling approaches exist, such as agent-based models (ABMs)~\cite{Muller2021ABM}, models based on partial differential equations (PDEs)~\cite{Foutel2022PDE}, and ordinary differential equation (ODE) models~\cite{Wulkow2021}. Each of these approaches offers distinct advantages: ABMs provide detailed individual-level dynamics but are computationally demanding; ODEs are efficient and analytically tractable but lack spatial structure; PDEs incorporate spatial variation, which is critical for modeling heterogeneous regions. 

Each of these modeling approaches has its place, but choosing the most appropriate one often depends on the characteristics of the domain being studied. In practice, however, many real-world regions exhibit a mix of features that make a single modeling paradigm insufficient. Consider, for instance, the Berlin-Brandenburg region: while Berlin represents a densely populated, relatively homogeneous urban area, Brandenburg is more rural and spatially heterogeneous. Applying the same model across both regions may either introduce unnecessary complexity or overlook important spatial effects. To address this, we adopt a hybrid modeling approach that assigns different models to distinct, non-overlapping subregions -- each selected to best capture the local dynamics. This strategy allows us to leverage the respective strengths of each modeling framework while maintaining consistency across the overall system.

Hybrid modeling approaches combining ABMs and PDEs have been employed in other scientific domains, such as cellular biology and oncology; however applications in the context of human-to-human disease transmission remain rare. For instance, Kostre et al.~\cite{Kostre2021Coupling} apply hybrid ABM-PDE models to simulate biochemical processes at the cellular level, modeling interactions between molecules and cells through coupled particle-based simulations and reaction-diffusion PDEs. Particles diffuse within the particle domain. Upon crossing the boundary into the concentration domain, they are eliminated. The concentration at the boundary of the PDE domain is converted into virtual particles and fractional virtual particles, which can jump into the particle domain at a certain rate. Similar ABM-PDE hybrids have been used to investigate viral dynamics in cell populations~\cite{Marzban2021hybrid}, antiviral drug effects~\cite{Juhász2024Probability}, and treatment strategies in oncology~\cite{Storey2021ABM}. 

Recent studies have also explored machine learning techniques to construct surrogate models that approximate behaviour of detailed ABMs~\cite{Claudio2022machine}. These data-driven approaches -- often based on neural networks -- can replicate complex ABM dynamics with significantly reduced computational cost, enabling more efficient parameter calibration and sensitivity analyses. As such, they offer a promising approach for enabling large-scale simulations. However, a key limitation of these methods lies in their lack of interpretability -- an important drawback in epidemiological applications, where understanding the underlying mechanisms is essential for informed public health decision-making.


Several studies have explored hybrid epidemic modeling approaches that combine agent-based representations with compartmental formulations (e.g.,~\cite{Zhao2023Joint,Singh2021Progression,Eden2021Agent}).
They develop frameworks that integrate agent-based network components with compartmental population-level structures for disease transmission, illustrating both the flexibility and, in certain contexts, the necessity of mixed modeling strategies in epidemiology. A similar principle underlies our Berlin–Brandenburg setting, where the choice of modeling paradigm is guided by structural heterogeneity. 

In this study, we present and evaluate a hybrid ABM-PDE model using real-world data, specifically applied to the Berlin-Brandenburg region. The model incorporates real-world mobile phone data, representing 25\% and 100\% of the population, capturing realistic movement patterns crucial for disease dynamics modeling. The model's evaluation is based on real-world infection data and compared to a high-resolution, full-ABM simulation that serves as a reference. 
Our hybrid approach enables the exchange of information between the two (sub-)models at each time step, ensuring a dynamic yet consistent interaction between local and global disease dynamics.


The resulting hybrid ABM-PDE model provides a faster alternative to a full-ABM model, offering the benefit of quicker parameter fitting due to reduced overall simulation runtime, regardless of whether the 25\% or 100\% population sample is used. Furthermore, the hybrid model demonstrated better accuracy over short simulation horizons -- approximately two months --, when fitting was done for two separate intervals. 
The underlying ABM is a streamlined version of EpiSim~\cite{GithubEpiSim} implemented in C++, and is capable of simulating 100\% of the population on a standard computer.

Building on these results, our long-term objective is to extend the current hybrid framework into a comprehensive ABM-PDE-ODE coupled model, which will integrate previous advancements in ABM-ODE~\cite{bostanci2025integrating} and PDE-ODE coupling~\cite{Kehrer2024Hybrid}. 

The complete model implementation and the processed infection data from the Robert Koch Institute (RKI) website are available on Zenodo~\cite{kehrer2025AbmPdeCode}. Mobility data is available from the authors upon reasonable request.

\section{Model Formulation} 

Our proposed hybrid model couples an agent-based model (ABM) and a partial differential equation (PDE) model to simulate the spatial spread of infectious diseases in the Berlin-Brandenburg region. The urban area of Berlin is represented by a PDE model that captures population-level dynamics and health state transitions, while the surrounding rural area of Brandenburg is modeled using an ABM based on real-world individual movement data from mobile phones. The data used for the ABM consists of individuals who have been in Berlin or Brandenburg at least once, with a few individuals from outside the region. For simplicity, we refer to the entire modelling domain as Brandenburg for the ABM contribution of the hybrid model and Berlin-Brandenburg for the full-ABM model. 

Both the ABM and the PDE model incorporate different epidemiological health states, and are dynamically coupled at each time step: individuals crossing the boundary are either removed from the ABM and added to the PDE as density contributions, or generated in the ABM from excess density in the PDE. This exchange ensures consistency across the boundary and reflects observed mobility patterns.

The remainder of this section is structured as follows:

\begin{itemize}
    \item Part~1 introduces the PDE-based compartment model for Berlin  (Section~\ref{sec:PDEModelDerivation}).
    \item Part~2 presents the trajectory-driven ABM for Brandenburg, based on mobile phone data (Section~\ref{sec:ABM}).
    \item The initialization -- or setting of initial conditions -- of both model component is described in Section~\ref{sec:InitialValues}.
    \item Finally, Section~\ref{sec:Coupling} explains the coupling mechanism that integrates both parts into a single hybrid simulation framework.
\end{itemize}

\subsection{Part 1: Berlin – PDE-Based Compartment Model} \label{sec:PDEModelDerivation}

To model disease dynamics in the urban region of Berlin, we derive a PDE system from a motion-based ABM that incorporates individual movement and includes health state transitions. This approach allows us to capture spatially continuous population dynamics while maintaining a structured epidemiological model. Several methods exist for deriving a (stochastic) PDE system from an ABM (with multiple states) (e.g.~\cite{Helfmann2021Interacting,Nardini2021Learning}). 
In our work, we follow the approach proposed by Helfmann et al.~\cite{Helfmann2021Interacting}, in which the authors show how to translate agent motion and state transitions into a system of stochastic PDEs. 

\begin{figure}[h!]
    \centering
    \includegraphics[width=0.4\textwidth]{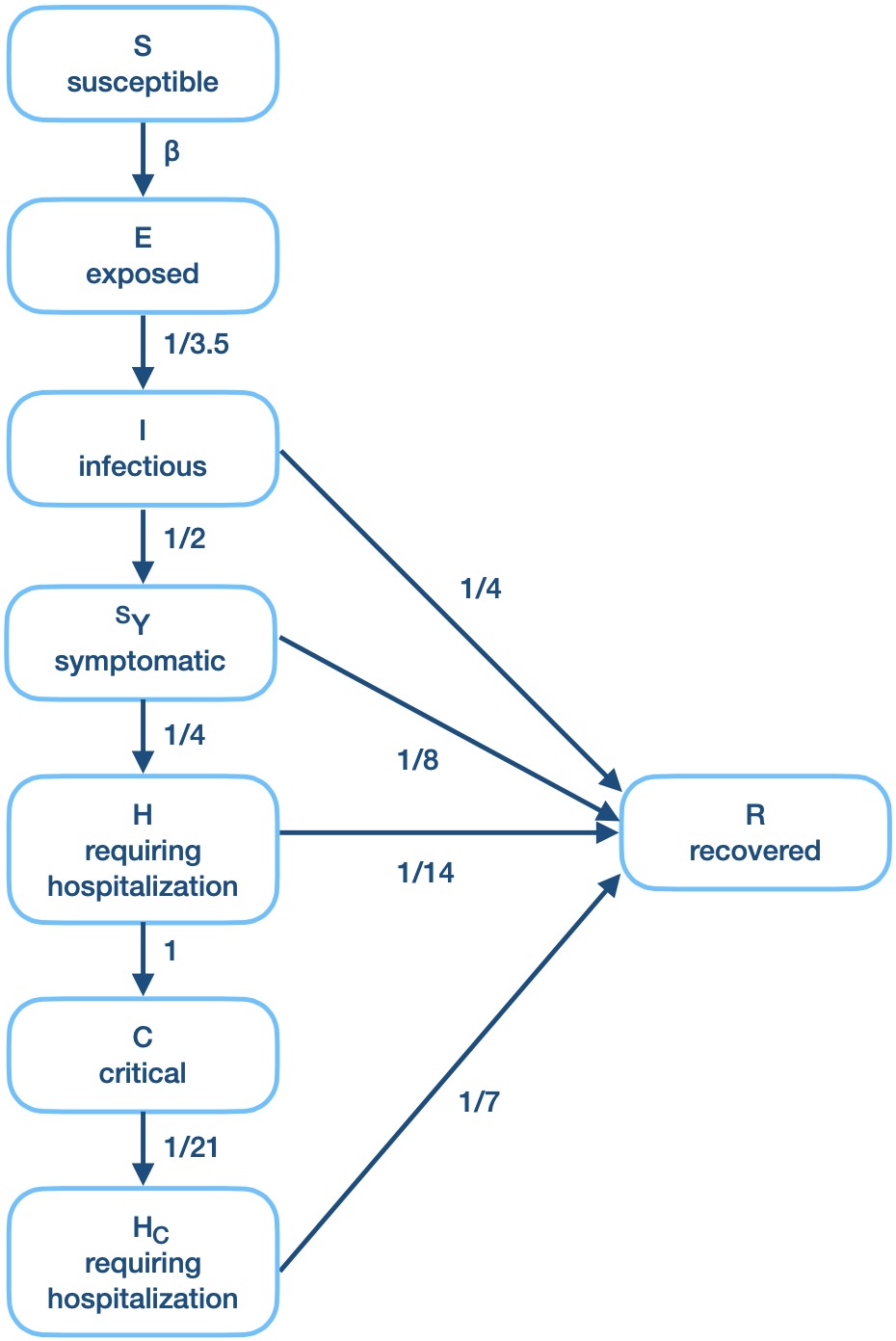}
    \caption{Structure of the SEIYHCR model with seven health states and ten transition rules, capturing disease progression from \textbf{S}usceptible to \textbf{R}ecovered via intermediate states.}
    \label{fig:compartments_flow}
\end{figure}

The health states in our model are: susceptible ($S$), exposed ($E$), infectious ($I$), symptomatic (\,$\sY$), requiring hospitalization ($H$ and $H_C$), critical ($C$), and recovered ($R$). The compartments and transition rates between compartments are displayed in Fig.~\ref{fig:compartments_flow}. The parameter choices are addressed in Section~\ref{sec:parameter-identification}.

\noindent
In the motion-based ABM, agents’ movement is governed by Brownian motion influenced by a landscape potential, which guides them toward frequently visited areas. State transitions occur at discrete time points with a fixed time step size and with a probability that depends on spatial proximity to other individuals and the corresponding transition rates. After reduction, each state variable $Y \in \{S, E, I, \sY, H, C, H_C, R\}$ evolves according to a stochastic reaction-diffusion equation of the general form:
\begin{align*}
    \frac{\partial Y(x,t)}{\partial t} 
    = D \Delta Y(x,t) + \divergence \left( \nabla V(x) \, Y(x,t) \right) + \text{(reaction terms)} + \text{(noise terms)}.
\end{align*}

Where -- as before -- $D$ is the diffusion coefficient, $V(x)$ is the landscape potential, and the reaction terms are derived from the interaction rules defined in the ABM (see Appendix (\ref{ABMrules})). The noise terms model stochastic fluctuations due to random movement and stochastic health status transitions at the agent level.

A full derivation, model specification, and the weak formulation used for numerical simulation are provided in the Appendix~\ref{appendix:PDEModelDerivation}.

\subsection{Part 2: Brandenburg – ABM Based on Mobile Phone Trajectories}
\label{sec:ABM}

In contrast to the PDE-based model for Berlin, the Brandenburg region is represented by an \textit{ABM} based on high-resolution mobile phone data to reconstruct realistic individual trajectories. Note that the used dataset includes individuals who have been present in Brandenburg or Berlin at least once in the recorded data, regardless of their actual place of residence. These individuals were kept in the simulation because they are part of the original data and contribute to the infection dynamics. Removing them would have created an artificial closed system and could have led to different transmission patterns. 

Each agent is either located inside a facility -- such as home, workplace, school or leisure facility -- or commuting between facilities. Each facility represents a specific location with an identifier and fixed spatial coordinates. It is assigned a category (e.g., home, school, workplace, leisure), which characterizes the interaction setting at that location. 
Within facilities, agents can only interact with others inside the same facility and category. This means, infection transmission can occur only if an infectious ($I$ or $\sY$) and a susceptible ($S$) agent are present in the same location of the same facility type. 
Health status updates happen at every time step, except for susceptible agents for which the health status is updated only upon leaving a facility. Consistent with the update mechanism of the underlying EpiSim framework, this modeling choice ensures that the infection process reflects accumulated exposure during the stay. Delaying the state update until departure prevents (multiple) updates within the same facility visit and avoids situations in which newly exposed individuals would immediately contribute to further transmission during that same stay. 
While commuting, agents are assumed to be isolated and cannot transmit infections. 
Previous implementations of EpiSim~\cite{Muller2021ABM,GithubEpiSim} allowed for transmission during commuting. However, in event-based models, transmission predominantly arises from prolonged contacts within facilities. While public transport is included as a potential transmission setting, it involves comparatively short contact durations and high air exchange rates. Therefore, commuting-related infections are not modeled.

The mobile phone data is provided as a sequence of time-stamped events for each agent. Each event indicates either the start or end of an activity at a specific facility of a certain category. Agents are assumed to travel in straight lines between facilities, with walking speed adjusted to match the expected arrival time. We do not explicitly model other modes of transportation. An agent’s health state (except for susceptible agents) may change during a time step that includes movement. This means a commuting agent may start a trip with one health state and enter the PDE domain with another. Changes in commuting patterns may alter the timing with which agents cross different modeling domains. Such shifts in timing can lead to earlier or later infection events and thereby influence the overall infection dynamics. 

Since facilities vary in size and layout, we estimate effective room sizes based on the maximum number of people present in the facility of a certain category during peak times 
(see Section \textit{Estimation of room sizes} in~\cite{Muller2020realistic,Muller2021ABM}). 
These estimates are used to approximate the density of interactions within facilities (and categories) and thus influence infection probabilities.


In addition to the baseline mobility extracted from the mobile phone dataset, the ABM incorporates mechanisms to simulate behavioral changes and policy interventions over time. One such mechanism is the use of activity change rates for out-of-home activities (see Sec.~\ref{sec:implementation}). These rates are defined on a daily basis and allow us to modify the original mobility data to reflect changes in social behavior, such as reduced movement during lockdowns. At the start of the simulation, the activity change rate is set to zero, meaning the input data is used as-is. On later days, if the rate becomes negative, a corresponding proportion of events (such as visits to work or leisure facilities) is randomly removed. This reduces contact opportunities while preserving the structure of the original data. If the rate becomes positive, we do not add new events, in order to avoid introducing artificial behavior not present in the original data. Policy interventions such as school closures are implemented in a similar way: if schools are marked as closed on a given day, all school-related events are excluded from the simulation for that period. %

The infection rate for a single susceptible individual in the ABM is determined by three components: (i) a baseline calibration parameter $\beta_{\text{const}}$, adopted from EpiSim and scaled by a constant factor to account for the simplified implementation used here, (ii) the contact intensity $ci$, and (iii) the cumulative interaction duration (measured in days) with infectious individuals during the time step. The contact intensity $ci$ depends on the maximum observed simultaneous occupancy of the respective facility (within its category), as derived from the mobility data, and on a category-specific infectivity term accounting for factors such as room size and air exchange conditions. Each facility category is assigned its own infectivity parameter. A detailed description of this formulation can be found in~\cite{Muller2021ABM}. All remaining transitions in health states are based on the rules described in Appendix~\ref{subsubsec:statusABM}.

\subsection{Initial Values} \label{sec:InitialValues}

Agents modeled in the ABM part of the hybrid system are initialized based on their first recorded positions from the mobile phone data. In the PDE component, the initial distribution of individuals is informed by the underlying landscape, which is derived from agent movement (see Appendix~\ref{appendix:PDEModelDerivation}). To use the landscape $V:\; \Omega_{Be} \rightarrow \mathbb{R}$ of domain Berlin $\Omega_{Be}\subseteq \mathbb{R}^2$ for initialization, we first shift all values so that the landscape becomes strictly positive, denoting the result as $V_+$. 
Next, we take the inverse of $V_+$, as locations with steeper landscape values correspond to areas where population density should be higher. We then normalize the inverse of $V_+$ to construct a probability distribution over the domain:
\begin{align} \label{eq:initial_distribution}
    \frac{1}{V_+(x_i) \int_{\Omega_{Be}} \frac{1}{V_+(x)} \; dx}.  
\end{align}
Multiplying this initial probability distribution by the total number of individuals in each health compartment yields the initial values of the PDE system as defined in equation (\ref{eq:PDE-model}).
%


\subsection{Coupling} 
\label{sec:Coupling}
%
%
The coupling between the ABM and PDE components is implemented by dynamically exchanging individuals between the two model domains. When agents cross the boundary from the ABM to the PDE region, they are removed from the ABM and their health status is translated into a corresponding density contribution in the PDE. Conversely, when individuals leave the PDE region, they are instantiated as new agents in the ABM, and the corresponding amount is subtracted from the PDE densities.

%
The coupling between the ABM and PDE components is implemented in a time-discrete manner and prescribed by the trajectory data. Within each time step, both submodels are first advanced independently according to their respective dynamics. The exchange of individuals between the two models is then performed explicitly before the next step begins. The transfer is realized as a redistribution of mass between an agent-based and a density-based representation. 

To ensure well-posedness of the PDE system and avoid artificial fluxes across the boundary, we impose zero Neumann boundary conditions for the PDE contribution (\ref{eq:PDE-model}) of the hybrid model across the entire boundary:
\begin{align*}
    D \nu^T \nabla Y = 0, && Y \in \{S,E,I,\sY,H,C,H_C,R\}
\end{align*}
where $\nu$ denotes the unit outer normal vector to the PDE domain $\Omega_{Be}$. 
Note that we still have agent-to-agent interaction within the PDE domain during each time step, as the agents will be deleted based on their location at the end of the time step. This means that they can still contribute to transmission dynamics within that time window.

The PDE domain is discretized using a finite element method on a triangular mesh (grid). The spatial domain is partitioned into triangles (triangulation), and the compartment densities are represented at the corresponding mesh (grid) nodes. In the following, the term “node” refers to a node of this triangulated mesh. A schematic illustration is provided in Appendix~\ref{appendix:grid} (see Fig.~\ref{fig:Berlin_grid_coarse}). 

From a numerical perspective, removing or adding individuals at a single grid node or their accumulation at the boundary through fluxes can lead to instabilities. Therefore, density adjustments are distributed proportionally over the domain according to the existing density distribution. This preserves positivity of the solution and results in stable dynamics. 

%
\subsubsection{Agent Enters PDE Domain}
When an agent leaves the ABM domain and enters the PDE domain, the agent is removed from the ABM and their contribution is added to the PDE model. 
The agent's health state is read and used to place them into the corresponding compartment of the PDE system. 
This procedure ensures consistent mass transfer across the model boundary and preserves the total population in the hybrid model. 

%
%
\subsubsection{PDE Person Enters Domain of ABM}
To simulate agents transitioning from the PDE domain (Berlin) into the ABM domain (Brandenburg), we precompute the expected number of individuals present in the PDE model for each hour of the day. If the time step corresponds to one hour, this allows us to determine how many individuals should leave the PDE model and be represented as agents in the ABM. For smaller time steps, we interpolate these values accordingly.

The health status of the transitioning individuals is not sampled stochastically but determined deterministically based on the current distribution of the PDE compartment densities. We compute the ratio of individuals to be removed from each compartment, proportional to their current densities. These initial ratios are adjusted so that we end up rounding down to whole individuals. Any remaining individuals are assigned to the susceptible compartment, which dominates the distribution. These ratios are then used to proportionally reduce the corresponding densities at each grid point, ensuring that more individuals are removed where the density is higher. 

To obtain the total number of individuals, we read the density values at each node and convert the density amount into the corresponding number of people. For this, we compute how much density is equivalent to a single person.
We assume that a single susceptible individual is to be added at a single grid point $x^*$, meaning that the density changes only at that specific location. This update is applied at the end of the time step, after the PDE solution has already been computed for all grid points $x$. We denote by $\tilde{S}(x,t_k)$ the PDE solution at time $t_k$ before adding the contribution from agent transitions. With these assumptions, the following equation is satisfied, ensuring that the total added density corresponds to one individual:
\begin{align*}
    \int_{\Omega_{Be}} S(x,t_k) - \tilde{S}(x,t_k) \, dx = 1,
\end{align*}
which is equivalent to
\begin{align*}
      \int_{\text{triangles touching }x^*} \varepsilon(x,t_k) \, dx = 1
\end{align*}
since all other triangles of the grid are not affected, and $\varepsilon$ denotes the density value equivalent to a single individual, which is to be determined. 
Further, we have
\begin{align*}
     \int_{\text{triangles touching }x^*} \varepsilon(x,t_k) \, dx &= \sum_{i=0}^{N_\triangle} \int_{\triangle_i} \varepsilon(x,t_k) \, dx 
     \\ & \approx \sum_{i=0}^{N_\triangle} \frac{\varepsilon(x^i_1,t_k) + \varepsilon(x^i_2,t_k) + \varepsilon(x^*,t_k)}{3}  \int_{\triangle_i} 1 \, dx,
\end{align*}
where $x^i_1$, $x^i_2$, $x^*$ denote the grid nodes of the triangle $\triangle_i$ and $N_\triangle$ is the total number of triangles touching the grid node $x^*$.
Since we are only changing the density value at grid node $x^*$, we have $\varepsilon(x^i_1,t_k)=\varepsilon(x^i_2,t_k)=0$ for $i=0,\dots,N_\triangle$. Hence, we obtain 
\begin{align*}
     \frac{ \varepsilon(x^*,t_k)}{3}  \sum_{i=0}^{N_\triangle}  |\triangle_i| \approx 1,
\end{align*}
which leads to 
\begin{align*}
     \varepsilon(x^*) \approx \frac{3}{\sum_{i=0}^{N_\triangle}  |\triangle_i|}.
\end{align*}

The event data contain complete daily trajectories for all individuals, and no new plans are generated during the simulation. A plan is considered “activated” when the corresponding individual enters the ABM domain. From that point on, the full mobility dataset is filtered so that only the events belonging to the agent IDs currently simulated in the ABM are processed. When an individual enters the ABM domain, we randomly pick a free agent ID, and the corresponding plan from the mobile phone-based event data is activated. This plan defines the agent's activities and locations for the current day. Occasionally, the agent's first or final position from that plan might be inside Berlin, i.e. outside of the ABM domain. We refer to such cases as a “less appropriate plan,” since the agent is then located in the PDE region although it was intended to remain within the ABM domain. These cases are not explicitly handled but occur infrequently in practice compared to the total population size. 
As agents enter Berlin during the timestep, they can still interact with other agents and contribute to the infection dynamics if a less appropriate plan was chosen, causing them to be placed in an unintended state. 

\section{Parameter Identification} \label{sec:parameter-identification}

Finding optimal parameters for a stochastic full-ABM or the ABM contribution of a hybrid ABM–PDE model presents significant challenges due to the computational cost and variability inherent in such simulations. Several approaches have been proposed to address this issue. For instance, one may reduce the number of runs~\cite{Muller2021ABM}, rely on more efficient surrogate models to speed up optimization~\cite{Helfmann2021Interacting,Perumal2025Surrogate,Claudio2022machine}, reformulate the existing model to make parameter inference more tractable~\cite{Lenti2024Variational}, adopt parameters from previous studies~\cite{Muller2020realistic,Muller2021ABM} or even rely on manual parameter estimation~\cite{Chakraborty2024brief}.
Surrogate approaches are often well suited for approximating fixed, aggregated outcomes of computationally expensive models over a predefined time horizon. However, accurately reproducing high-dimensional time-dependent dynamics represents a substantially more demanding task. 
In EpiSim, for example, parameters are fitted by minimizing the Root Mean Squared Logarithmic Error 
between the simulated and observed number of hospitalized cases~\cite{Muller2021ABM}. Here, the authors use multiple (10 or 30) Monte Carlo runs and average the results to obtain stable estimates based on eight independent Monte Carlo seeds~\cite{Muller2021ABM}. 

In our model, we used the previously optimized calibration parameter $\beta_{\text{const}}$ for the infection rate for Berlin from the original EpiSim (see \textit{BerlinSensitivityRuns.java} in~\cite{GithubEpiSim}), adjusted by a factor to account for the changes in our implementation (see Sec.~\ref{sec:implementation}). The remaining parameters were taken from existing literature~\cite{Muller2020realistic,Muller2021ABM}. For the Brandenburg model, we apply the same rates, since no region-specific calibration was available and the overall accuracy was sufficient for our focus on evaluating the coupling approach.

In contrast to the ABM, implementing the infection dynamics in the PDE model is more difficult, as the ABM's infection process depends on numerous factors  
-- such as shedding and intake rates, duration of interaction with infected agents, contact intensity, a calibration parameter corresponding to the infection rate in a compartmental model~\cite{Muller2020realistic}, and daily activity change rates -- where some of these factors vary over time and across different facility categories. Translating these effects directly to the PDE model is challenging. One possibility is to only have the independent infection rate, which is constant with respect to time and fit it to target data.  

Instead, we incorporate data that is applicable to all agents and for all categories, i.e., the activity change rates. For an ABM, implementing mild restrictions by considering changes in activity participation would alter agent trajectories and, consequently, the overall landscape. However, this change in landscape would likely not reduce infections but merely shift their locations. 
The implementation of activity reductions has the intended effect of reducing infection numbers, resulting, e.g., in a lower infection rate in the PDE system. 
For this reason, we directly incorporated changes in activity participation into the infection rate. 
We assume that a reduction in out-of-home activities corresponds to a decrease in infection numbers, and thus in the infection rate. We define the infection rate of the PDE model (\ref{eq:PDE-model}) as
\begin{align*}
    \beta = \Biggl(1- \frac{\text{out-of-home activity change rate in \%}}{100\%} \Biggr) \frac{\beta_{\text{const}} }{\pi r^2},
\end{align*}
where $\beta_{\text{const}}$ is the constant part of the infection rate that will be used for fitting and $r$ is the contact radius (see Appendix~\ref{sec:weakformulation}). 
Since the contact radius is unknown, we determine it indirectly through the fitting of $\beta_{\text{const}}$, given that
\begin{align*}
    \beta \pi r^2 = \Biggl(1- \frac{\text{out-of-home activity change rate in \%}}{100\%} \Biggr) \beta_{\text{const}}.
\end{align*}
For the ABM, we have the calibration parameter $\beta_{\text{const}}$ corresponding to the infection rate in a compartmental model~\cite{Muller2020realistic}. All final transition rates can be found in the next Section~\ref{sec:implementation} in Tables~\ref{table:params} and~\ref{table:params_beta}.


School closures are handled differently, as they typically occur over a clearly defined time interval, leading to a more pronounced impact. Rather than scaling the infection rate or adjusting it to the changed landscape, we divide the simulation time interval into two phases -- one without school closures and one with school closures -- fitting the infection rates separately for these subintervals. In contrast, accounting for daily fluctuations in activity changes in the form of mild restrictions would require fitting infection rates separately for each day, which is not feasible. Therefore, we directly incorporated these changes into the infection rate to account for their effects.
Note that the constant part of the infection rate $\beta_{\text{const}}$ is fitted separately for the PDE and ABM parts within the hybrid model, as their dynamics differ. The fitting is done using a simple grid search (see Sec.~\ref{sec:implementation}). 

All other transition parameters are assumed to be the same across the ABM and PDE model. This is justified by the similarity of our trajectory-based ABM and the motion-based ABM (with landscape, see Appendix~\ref{subsubsec:motionABM}) from which the PDE is derived~\cite{Helfmann2021Interacting}. If Brandenburg were also modeled using PDEs, the same infection rate as in Berlin would be applied, consistent with the ABM design, where a single infection rate governs the entire spatial domain.

In principle, an ABM with simplified infection dynamics -- and thus a correspondingly simplified hybrid model -- could even serve as a reduced model for parameter inference, replacing the full-ABM in optimization loops. However, our focus in this paper lies not on optimal parameter fitting but on developing and evaluating a coupling mechanism between ABM and PDE approaches that can be applied to real-world data. 

\section{Implementation} \label{sec:implementation}

The available mobile phone dataset contains event data for three distinct day types: weekdays, Saturday, and Sunday. This allows us to differentiate between typical mobility patterns across the week. We start our simulation on March 2, 2020, which is a Monday, and thus begin with weekday data. Agents are allowed to move beyond Brandenburg and Berlin, as the mobile phone data captures their locations across a wider region. These event-based trajectories are time-stamped and specify when an individual enters or leaves a facility of a certain type (e.g., home, work, leisure), allowing us to reconstruct realistic daily schedules.

To facilitate the exchange between ABM and PDE in our hybrid model, the event data from the mobility dataset are sliced into chunks corresponding to individual time steps. For each step, the function \texttt{filter\_event\_data} verifies whether events exist exactly at the start and end. Otherwise, it identifies the events surrounding the start and end of the time step, from which artificial boundary events are reconstructed. If no event is recorded before the beginning of a time step (e.g., 4 a.m.), the final events of the previous day are inspected. Typically, individuals remained at home overnight, so the last known location is propagated forward. Artificial events, together with all events occurring in between, are transferred to a sliced data matrix, restricting processing to the relevant time step. Artificial events are used to track the last known positions of agents, allowing efficient initiation of the transition process, and are also used in the hybrid model for visualization purposes. For simplicity, we applied the same slicing procedure to the full-ABM, which may introduce some computational overhead.

\begin{figure}[h]
    \centering
    \includegraphics[width=0.5\linewidth]{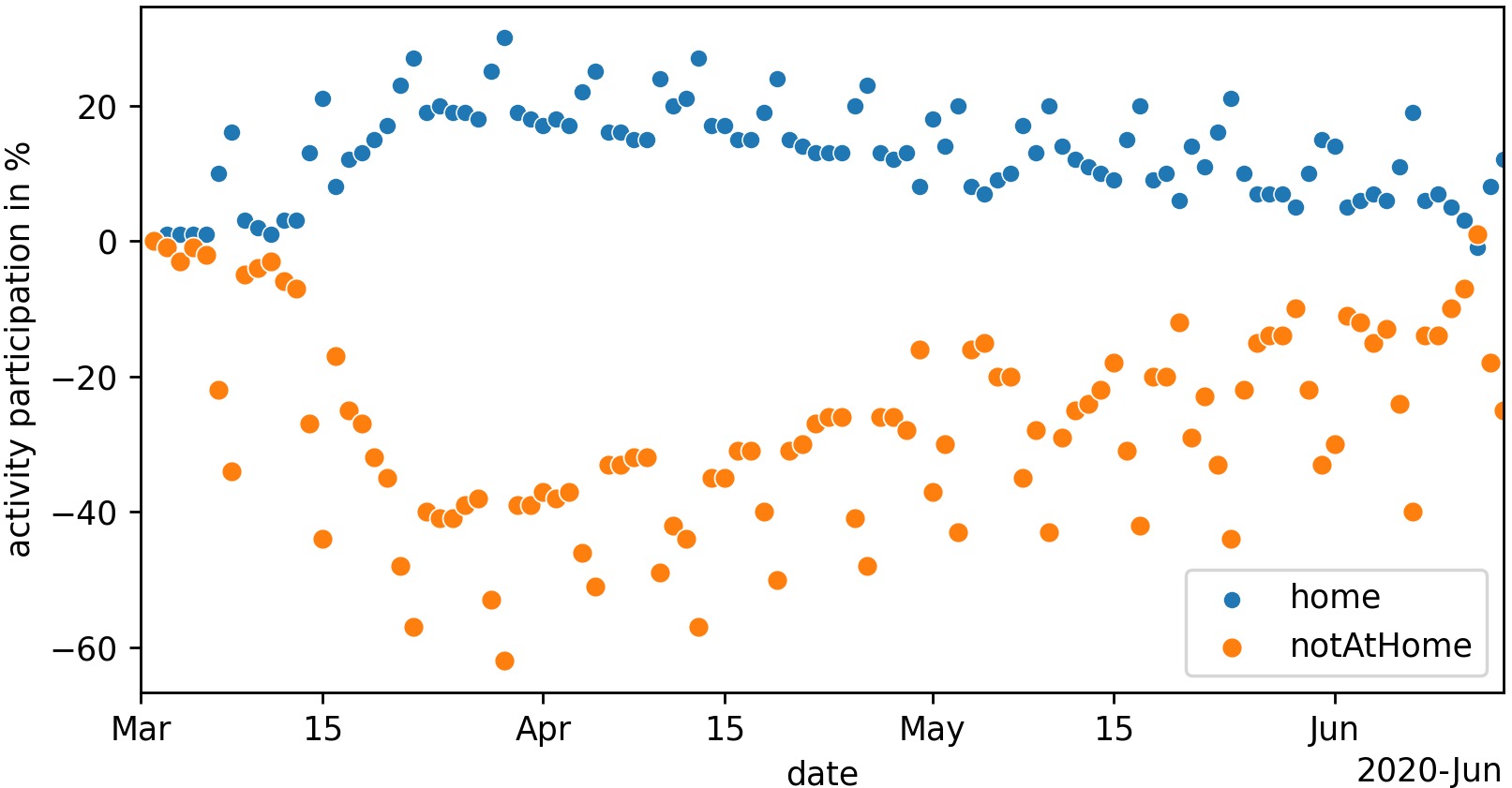}
    \caption{Activity participation in \% for activities at home and not at home including weekends.}
    \label{fig:activityChanges}
\end{figure}

In addition, we incorporate activity change rates to account for behavioral adaptations over time, such as reductions in out-of-home activities during the early phases of the pandemic. These rates also begin on March 2, 2020 (which matches the data provided in \textit{be\_2020\-mobility\_data.csv}~\cite{balmer2022Synthetic}), covering the period until June 12, 2020. The activity reductions are visualized in Figure~\ref{fig:activityChanges} and are used to modulate infection dynamics in both the ABM and PDE components, as described in Section~\ref{sec:ABM} for the ABM and in Section~\ref{sec:parameter-identification} for the PDE model. The first day assumes no reduction in activity participation and serves as a baseline. 

The activity change rates represent relatively mild, day-specific restrictions in behavior. More substantial interventions -- such as school closures -- are considered starting from March 16, 2020. For simplicity and consistency in the simulation, we assume the same school closure date for both Berlin and Brandenburg, even though the actual dates differed slightly. According to~\cite{Muller2021ABM}, school closures in Berlin were implemented on March 16, 2020, while~\cite{wikipedia_covid19_berlin} mentions that most schools were closed on March 16, but the official closure was on March 17, 2020. In Brandenburg,~\cite{wikipedia_covid19_brandenburg} states that schools were officially closed on March 18, 2020.


Although mask-wearing has been implemented in the code, it was not required for the conducted experiments and thus has no effect on the presented results. In Berlin, mask mandates were introduced earlier in public transport (from April 27, 2020) and subsequently extended to public spaces such as shops on April 29, 2020~\cite{wikipedia_covid19_berlin}. Since our model does not account for transmission in public transport, we defined April 28, 2020 as the final day of the simulation. From this point onward, a new modeling interval and a recalibration of the PDE infection rate would have been necessary. 

To initialize the population distribution, we used the previously defined landscape (see Eq.~(\ref{eq:landscape_definition})) to derive the initial distribution (see Eq.~(\ref{eq:initial_distribution})). We visualized the process of landscape creation in the Appendix~\ref{appendix:landscapeCreation}. 
The landscape, as defined in equation (\ref{eq:landscape_definition}), depends on the diffusion coefficient. We computed the landscape and its gradient once and rescaled it according to changes in the diffusion coefficient. 
The gradient of the landscape is computed by first interpolating it from a rectangular grid (derived from the histogram of agent trajectories) to a triangular grid (used in the PDE model) using the \texttt{RegularGridInterpolator} from \texttt{scipy.interpolate}. The landscape is then defined for every grid node on the triangular grid. By incorporating information from the triangulation, which determines how grid points are connected by triangular elements, we create an object of the class \texttt{Triangulation} using \texttt{matplotlib.tri} and access its gradient using the \texttt{gradient} method.

The diffusion coefficient was selected to ensure limited individual movement over the short simulation period. We chose a value of $1e-6$. 
The time step for simulation was set to half an hour ($\nicefrac{1}{48}$ days), as we observed a significant improvement in accuracy compared to one-hour steps, with little further gain from finer time discretization. 

For fitting the constant part of the infection rate in the PDE model (\ref{eq:PDE-model}), we used a simple grid search. We compared different parameter values by computing the mean absolute error (MAE) between the mean of 20 simulation runs and the target data. Before computing the MAE, both the simulation results and the target data were scaled to the full population size to ensure comparability of the reported outcomes. For the first time interval (March 2, 2020, to March 15, 2020), we initially tested $\beta_{\text{const}} \in \{4e+2,4.5e+2,5e+2,5.5e+2\}$, while for the second time interval (March 16, 2020, to April 28, 2020), we considered $\beta_{\text{const}} \in \{1.2e+2,1.3e+2,1.4e+2,1.5e+2,1.6e+2,1.7e+2,1.8e+2,1.9e+2\}$ (see Table~\ref{table:grid_search_PDE}). 
We tested all combinations together, as the best choice for the first interval may minimize the error in that interval but could be far off for the second interval. The second interval begins with the implementation of school closures and is characterized by a significant change in the trajectory of the infection curve.
The smallest average error was achieved with the parameters $4.5e+2$ and $1.6e+2$ (see Table~\ref{table:grid_search_PDE}). For the full population, which is four times larger than the partial population, these parameters can be scaled accordingly by a factor of $\nicefrac{1}{4}$. 
The parameter fitting for the scalar factor of the ABM infection rate was also performed using a simple grid search (see Tables~\ref{table:grid_search_ABM_25pt} and~\ref{table:grid_search_ABM_100pt}). All final transition rates can be found in Tables~\ref{table:params} and~\ref{table:params_beta}.

It is important to emphasize that the grid search was designed to be fast and coarse, sufficient for demonstrating the functionality of the hybrid model. Although the infection rate could be further optimized using more refined step sizes or advanced optimization methods, this was not the focus of our work. Since slight variations in these parameters do not significantly impact our overall conclusions, a more detailed optimization is beyond the scope of this paper. 

\begin{table}[h!]
    \centering
    \begin{tabular}{cc||cccc}
     & & \multicolumn{4}{c}{March 2, 2020 - March 15, 2020} \\
    \cline{3-6}
             & &  $4e+2$  &  $4.5e+2$  & $5e+2$  & $5.5e+2$  
             \\
    \hline \hline
\multirow{8}{*}{\rotatebox{90}{\shortstack{March 16, - \\ April 28, 2020}}} & 
      \multicolumn{1}{|c||}{$1.2e+2$}  & $90.14$  & $55.93$  & $42.32$ & $61.07$ 
      \\
&     \multicolumn{1}{|c||}{$1.3e+2$}  & $86.69$ & $48.38$ & $38.23$ & $51.25$  
\\
&     \multicolumn{1}{|c||}{$1.4e+2$}  & $67.16$ & $45.48$ & $46.16$ & $66.83$ 
\\
&     \multicolumn{1}{|c||}{$1.5e+2$}  & $65.22$ & $28.75$ & $49.13$ & $113.77$ 
\\
&     \multicolumn{1}{|c||}{$1.6e+2$}  & $52.36$ & \textbf{25.12} & $69.02$ & $138.23$ 
\\
&     \multicolumn{1}{|c||}{$1.7e+2$}  & $52.24$ & $60.46$ & $92.07$ & $198.72$ 
\\
&     \multicolumn{1}{|c||}{$1.8e+2$}  & $43.54$ & $55.96$ & $115.88$ & $196.12$ 
\\
&     \multicolumn{1}{|c||}{$1.9e+2$}  & $49.32$ & $82.53$ & $154.87$ & $260.46$ 
\end{tabular}
    \caption{Mean absolute error of the grid search (20 runs) for the hybrid model using a 25\% population sample. The smallest error was achieved with the parameters $4.5e+2$ and $1.6e+2$. } 
    \label{table:grid_search_PDE}
\end{table}
%
\begin{table}[h!]
    \centering
    \begin{tabular}{c||c|c|c|c|c|c}
& \multicolumn{6}{c}{March 2, 2020 - April 28, 2020} \\
\cline{2-7}
& \multicolumn{6}{c}{ABM calibration parameter factor $ \cdot 1.7e-5$ } \\
\hline \hline
factor & $0.25$ & $0.3$ & $0.35$ & $0.4$ & $0.45$ & $0.5$ \\
MAE & $126.06$ & $113.53$ & $82.06$ & \textbf{27.91} & $83.94$ & $266.34$
\end{tabular}
    \caption{Mean absolute error (MAE) of the grid search (20 runs) for the full-ABM using a 25\% population sample. The smallest error was achieved with factor $0.4$.}
    \label{table:grid_search_ABM_25pt}
\end{table}

%
\begin{table}[h!]
    \centering
    \begin{tabular}{c||c|c|c|c|c|c}
& \multicolumn{6}{c}{March 2, 2020 - April 28, 2020} \\
\cline{2-7}
& \multicolumn{6}{c}{ABM calibration parameter factor $ \cdot 1.7e-5$} \\
\hline \hline
factor & $0.25$ & $0.3$ & $0.35$ & $0.4$ & $0.45$ & $0.5$ \\
MAE &  $127.63$ & $100.2$ &    \textbf{63.77}  &   $148.42$       &  $291.9$       &    $931.47$
\end{tabular}
    \caption{Mean absolute error (MAE) of the grid search (20 runs) for the full-ABM using a 100\% population sample. The smallest error was achieved with factor $0.35$.}
    \label{table:grid_search_ABM_100pt}
\end{table}
\begin{table}[h]
    \centering
    \begin{tabular}{c|c||c|c}
       &  & from March 2, 2020 & from March 16, 2020 \\
        \hline \hline
      \multirow{2}{*}{25\%} & full-ABM  &  $0.4 \cdot 1.7e-5$ & $0.4 \cdot 1.7e-5$ \\
      & hybrid & $4.5e+2$ & $1.6e+2$ \\
      \hline
      \multirow{2}{*}{100\%} & full-ABM  &  $0.35 \cdot 1.7e-5$ & $0.35 \cdot 1.7e-5$ \\
      & hybrid & $1.125e+2$ & $4e+1$ 
    \end{tabular} 
    \caption{Constant part of the infection rate $\beta_{\text{const}}$ of the full-ABM and hybrid model for both 25\% and 100\% population samples.}
    \label{table:params_beta}
\end{table} 
\begin{table}[h]
    \centering
    \begin{tabular}{c||ccccccccc}
     & \multicolumn{9}{c}{March 2, 2020 - April 28, 2020} \\
        \cline{2-10}
         & $\sigma$ & $\gamma$ & $\eta$  & $\kappa$ & $\eta_c$  & $\phi_i$ & $\phi_{sy}$ & $\phi_h$ & $\phi_{hc}$  \\
        \hline \hline
      full-ABM  & $\nicefrac{1}{3.5}$ & $\nicefrac{1}{2}$ & $\nicefrac{1}{4}$ & $1$ &  $\nicefrac{1}{21}$ & $\nicefrac{1}{4}$ & $\nicefrac{1}{8}$ & $\nicefrac{1}{14}$ &  $\nicefrac{1}{7}$ \\
      hybrid & $\nicefrac{1}{3.5}$ & $\nicefrac{1}{2}$ & $\nicefrac{1}{4}$ & $1$ & $\nicefrac{1}{21}$ & $\nicefrac{1}{4}$ & $\nicefrac{1}{8}$ & $\nicefrac{1}{14}$ &  $\nicefrac{1}{7}$ 
    \end{tabular} 
    \caption{Transition rates (excluding infection rate) of the full-ABM and hybrid model for both 25\% and 100\% population samples.}
    \label{table:params}
\end{table} 
The density values are checked to ensure they are non-negative at least once for each time step.

We created the modeling domain of Berlin using the boundary grid nodes of OpenDataLab~\cite{RandwerteDaten}. The triangulation of the modeling domain Berlin was performed using Triangle~\cite{triangle}, while the hybrid ABM-PDE systems were solved using the finite element method implemented in the Kaskade7 software~\cite{GoetschelSchielaWeiser2020}, with the Dune interface. 
The visualization of our simulation results was done using the \texttt{matplotlib} package in Python 3.9 and the grid was visualized using ParaView~\cite{paraview}. 

We computed the 7-day average of Covid-19 data from the RKI~\cite{rki2025SARS} 
and used it as target data for parameter fitting, evaluation, and to define the initial conditions of the PDE system. 
For the ABM, the data were scaled by a factor of 20. Due to stochasticity in the ABM, infections can die out more easily in the early phase. Therefore, higher initial values were chosen compared to the official RKI data. Larger scaling factors were tested but did not lead to further qualitative changes, as extinction events no longer occurred.

Our ABM is a streamlined version of EpiSim~\cite{GithubEpiSim}, implemented following\cite{Muller2020realistic,Muller2021ABM}, with the exclusion of infection dynamics outside of facilities and without the use of weather data. 

For computational reasons, previous studies such as~\cite{Muller2020realistic} used a 25\% sample of the full population and reported all results after scaling them to 100\%. The 25\% mobility dataset for Berlin is publicly available~\cite{balmer2022Synthetic}. In our study, we do not use the Berlin data alone but rather the combined data for Brandenburg and Berlin, where the Berlin data is a subset of our dataset. It can be modified for independent tests or used to understand the data structure.

While prior work indicates that the outcomes of the 25\% and 100\% population models are largely similar, we explicitly examine both in this paper. 
The full dataset used in our simulation consists of 1,497,966 individuals for the 25\% population sample and 6,233,887 individuals for the 100\% population sample. Of these, 893,927 and 3,576,121 individuals are located in Berlin, and 604,039 and 2,657,766 in Brandenburg or other parts of Germany included in the dataset, respectively. 
We measure both the average error and the average runtime to better understand the performance variations between the models. 
Particularly for 
the PDE contribution of the hybrid model, we aim to assess how well the rates from the 25\% model can be transferred to the 100\% model. 
This will help determine whether parameter optimization can be performed using faster models and, for example, enable predictions for the entire population. Additionally, it will be interesting to compare whether the hybrid model with 100\% of the population or the ABM with 25\% of the population is faster, and which model potentially results in a lower error. 

Implementing the model for 100\% of the population results in changes to several data files. The event data file will be updated to include more people with events, the facility data file will expand to list additional facilities, and the maximum number of agents per facility in a given category will increase as more agents visit the same locations. Consequently, more facilities will be included overall. In~\cite{Muller2020realistic,Muller2021ABM}, it was mentioned that $N^\text{spacesPerFacility}$ needs to be adjusted for the entire population. However, we will not modify this parameter, as other factors, such as room size, will also change, effectively counteracting the need to scale $N^\text{spacesPerFacility}$. The calibration parameter $\beta_{\text{const}}$ will be optimized. 

It is worth noting that the infection rate in the ABM depends on multiple factors. Among them, contact intensity is probably the most affected by scaling, since the number of agents per facility changes with dataset size. The infection rate $\beta$ in the PDE model, however, must be explicitly rescaled. This is due to its nonlinear structure: it includes terms that represent pairwise interactions (e.g., between susceptibles and infectious), which scale quadratically with population size. In contrast, the other transition rates represent individual-level processes and remain unchanged.

While infections in the simulation occur across various locations and times, official case data reported by the Robert Koch Institute (RKI) are aggregated by the main residence of infected individuals, in accordance with \S~11 of the Infection Protection Act (IfSG)~\cite{IfSG11}. Therefore, to match simulation results with reported data, we sum the number of symptomatic individuals in the simulation across all locations. However, this may lead to overestimation, as our model includes individuals who live outside Berlin and Brandenburg but were present in the simulation domain.

Tracking the health status of each agent across simulation days requires identifying agent IDs. This is feasible in the ABM but not in the PDE contribution of the hybrid model, where no agent IDs exist. As a result, some information is inevitably lost in the hybrid setting. For instance, if a recovered individual enters the PDE domain and later reappears in the ABM, they may be reinitialized as susceptible. This does not align with our health state transitions and highlights a key limitation of our hybrid model (see Appendix~\ref{subsubsec:statusABM}).



%
%
\section{Numerical Experiments}
In this section, we present the numerical results of our simulations for both the pure ABM and the hybrid model, which integrates both agent-based and PDE-based components. 
For readers interested in the technical details underlying these simulations, additional information is provided in the appendix. This includes the construction and key properties of the PDE grid (see Appendix~\ref{appendix:grid}), the process of landscape creation used to initialize the population distribution (see Appendix~\ref{appendix:landscapeCreation}), and the number of simulation runs performed, along with performance evaluation in terms of runtime and error metrics (see Appendix~\ref{appendix:NumberOfRuns}). In Section~\ref{subsec:compTimeAccuracy}, we compare the computational efficiency and accuracy of the ABM and hybrid model, highlighting the trade-offs introduced by coupling the two approaches. Finally, Section~\ref{subsec:extremeExperiments} presents a set of extreme-case experiments designed to test the robustness and behavior of the hybrid model under challenging conditions. All experiments are conducted for two different population scenarios: a 25\% sample of the full dataset, with results scaled accordingly, and the complete 100\% population. 

\subsection{Computational Time vs. Accuracy} \label{subsec:compTimeAccuracy}
Next, we will compare computational time and accuracy of both models for both population sizes. 
Although Appendix~\ref{appendix:NumberOfRuns} demonstrated that the hybrid model requires fewer runs than the full-ABM for the same threshold, we still evaluate 100 runs for all models. Since these runs were already generated in the previous experiment, we simply reuse them for this analysis. We assess accuracy based on the mean and standard deviation over all runs, with values plotted up to the final time $T = 58$, while the overall mean across the entire time period for all runs is reported. Additionally, we measure the change in accuracy and computational time from the full-ABM to the hybrid model using the relative difference in their mean values, given by
\begin{align*}
    \frac{\mu_{\text{hybrid}}-\mu_{\text{full-ABM}}}{\mu_{\text{full-ABM}}},
\end{align*}
where $\mu_{\text{hybrid}}$ and $\mu_{\text{full-ABM}}$ denote the (absolute) mean values of the hybrid and full-ABM models, respectively. 
Computational time was measured with the parameter setting \texttt{verbosity=0} in the \texttt{covid.cpp} file, which reduces print outputs, prevents the creation of plots, disables the saving of agent health state and location data, and avoids storing density-related information. The parameter setting \texttt{verbosity=1} prints and stores all information except agent details.

\subsubsection{25\% Population}
\begin{figure}[h!]
    \centering
    \begin{subfigure}[b]{0.49\textwidth}
        \centering
        \includegraphics[width=\textwidth]{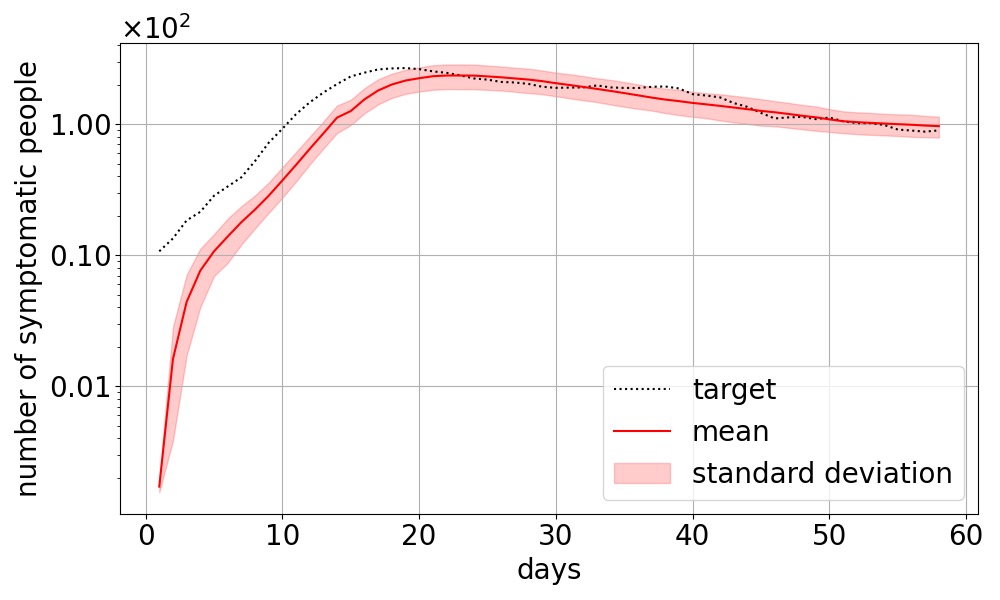}
        \caption{Hybrid model.}
        \label{fig:results_hybrid_25pt}
    \end{subfigure}
    \hfill
    \begin{subfigure}[b]{0.49\textwidth}
        \centering
        \includegraphics[width=\textwidth]{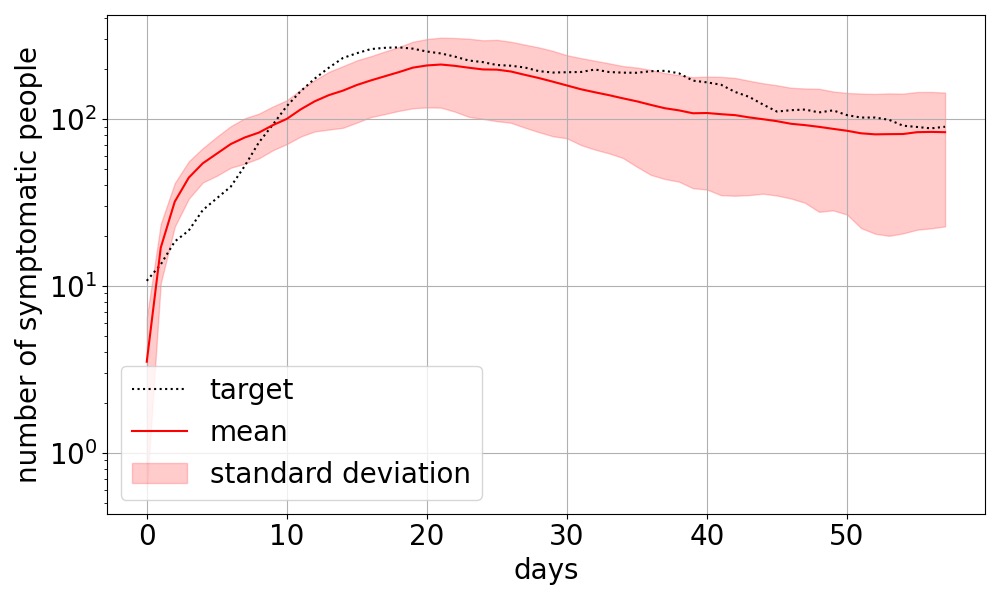}
        \caption{Full-ABM.}
        \label{fig:results_full_ABM_25pt}
    \end{subfigure}
    \caption{Mean and standard deviation of the number of symptomatic individuals and target values for a 25\% population sample in Berlin-Brandenburg (100 runs).}
\end{figure}

The fitted results of the hybrid model and the full-ABM, along with their target data, can be observed in Figs.~\ref{fig:results_hybrid_25pt} and~\ref{fig:results_full_ABM_25pt}, respectively. Note that for the 25\% population experiments, the reported number of symptomatic individuals is scaled by a factor of four to represent the full population. 
The results represent the mean over 100 runs, with the corresponding standard deviation shown in the figures.
The results for the error of the full-ABM and hybrid model are shown in Table~\ref{table:error_25pt}, while the duration for both models is presented in Table~\ref{table:duration_25pt}. Unlike in the original paper~\cite{Muller2021ABM}, the error was not measured logarithmically. The error of the hybrid model is likely lower than that of the ABM due to the separate fitting of the two time intervals. However, this does not guarantee that the same holds for other federal states or subregions in general, nor for longer time intervals. Furthermore, for the current hybrid model and its implementation, there is no guarantee that the error will remain lower if the time interval is not split. It is also important to note that the parameter optimization was conducted in a coarse manner, and better results could potentially be achieved for both the ABM and the hybrid model.

Since the infection dynamics in the PDE model are mainly driven by activity data and fewer direct data are used compared to the ABM, the PDE model -- and thus the hybrid model -- may be less suitable for predictions when compared to the ABM. However, the advantage of the hybrid model with its lower error is that we achieved a very good fit using the same compartments as the ABM, but with less complex infection dynamics, all while conducting a fast and rough parameter fitting.
\begin{table}[h!]
    \centering
    \begin{tabular}{c|cc}
         & full-ABM & hybrid \\
         \hline
       absolute mean error to target & 
       $35.37$ & $25.73$
       \\ 
        increased error to previous model & - & 
        $-27.25$\%
    \end{tabular}
    \caption{Error of the full-ABM and hybrid model (100 runs) using a 25\% population sample.}
    \label{table:error_25pt}
\end{table}
\begin{table}[h!]
    \centering
    \begin{tabular}{c|cc}
         & full-ABM & hybrid \\
         \hline
        mean duration in min & 
        $40.46$ & $31.62$ \\
        decreased duration to previous model & - & 
        $21.85$\%
    \end{tabular}
    \caption{Duration of the full-ABM and hybrid model (10 runs) using a 25\% population sample.}
    \label{table:duration_25pt}
\end{table}

\subsubsection{100\% Population}
\begin{figure}[h!]
    \centering
    \begin{subfigure}[b]{0.49\textwidth}
        \centering
        \includegraphics[width=\textwidth]{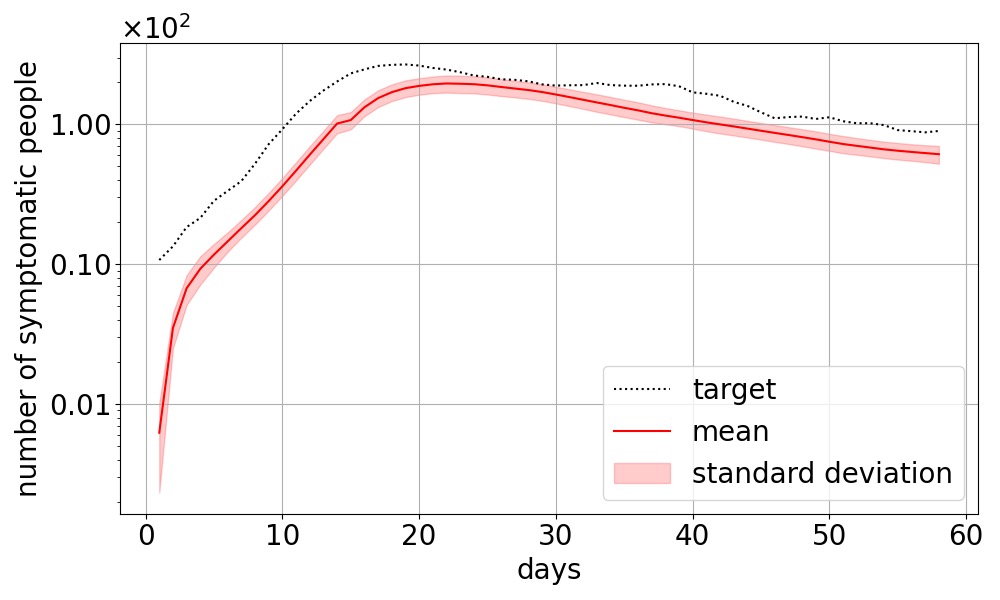}
        \caption{Hybrid model.}
        \label{fig:results_hybrid_100pt}
    \end{subfigure}
    \hfill
    \begin{subfigure}[b]{0.49\textwidth}
        \centering
        \includegraphics[width=\textwidth]{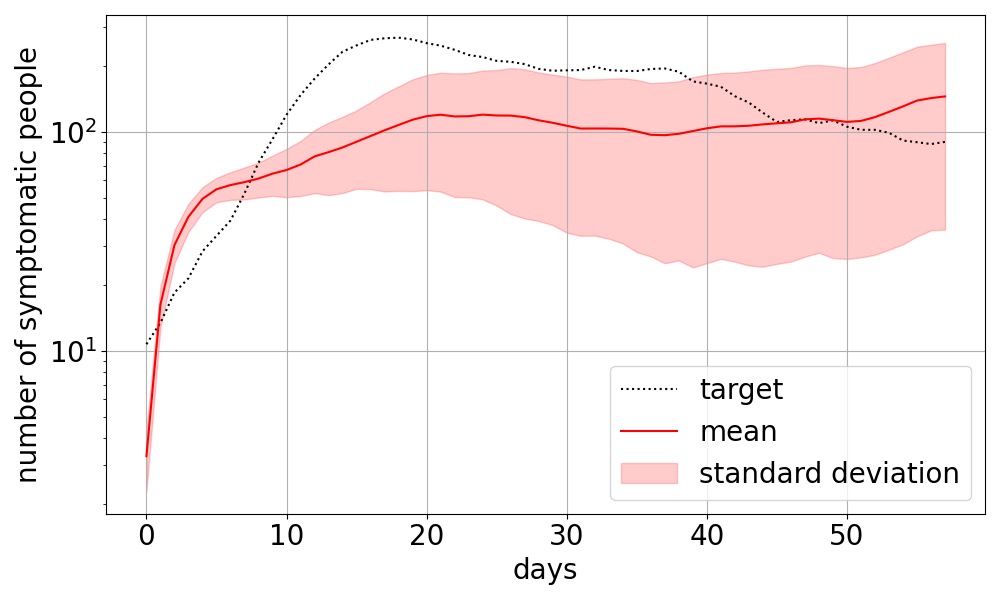}
        \caption{Full-ABM.}
        \label{fig:results_full_ABM_100pt}
    \end{subfigure}
    \caption{Mean and standard deviation of the number of symptomatic individuals and target values for the full population sample (100\%) in Berlin-Brandenburg (100 runs).}
\end{figure}

The fitted results for the hybrid model and the full-ABM, along with their corresponding target data, are shown in Figs.~\ref{fig:results_hybrid_100pt} and~\ref{fig:results_full_ABM_100pt}, respectively. Again, these results represent the mean across 100 simulation runs, with the associated standard deviation displayed in the figures. In contrast to the 25\% scenario (where results were scaled to 100\%), the results correspond to simulations performed directly on the full population. 
The error results for the full-ABM and hybrid model are presented in Table~\ref{table:error_100pt}, while the runtime for both models is shown in Table~\ref{table:duration_100pt}. The fitting was performed using the 25\% population dataset for the PDE contribution of the hybrid model, so we already expected higher errors when applying it to the full population dataset. This is indeed the case: the error almost doubled. However, we did not expect the simulation results of the full-ABM for the total population, obtained using fitted parameters, to exhibit an error almost twice as large as those of the model for the partial population (25\%) using its corresponding fitted parameters. The increased error compared to the previous model went from -27.25\% to -27.44\%.

The runtime of the hybrid model and the ABM have increased by almost a factor of five. As a result, the decreased duration compared to the previous model rises from 21.85\% to 23.49\%.
\begin{table}[h!]
    \centering
    \begin{tabular}{c|cc}
         & full-ABM & hybrid \\
         \hline
       absolute mean error to target & 
       $65.49$ & $47.52$   \\ 
        increased error to previous model & - & 
        $-27.44$\%
    \end{tabular}
    \caption{Error of the full-ABM and hybrid model (100 runs) using the full population sample (100\%).}
    \label{table:error_100pt}
\end{table}
\begin{table}[h!]
    \centering
    \begin{tabular}{c|cc} 
         & full-ABM & hybrid \\
         \hline
        mean duration in min & 
        $197.41$ & $151.03$ \\
        decreased duration to previous model & - & 
        $23.49$\%
    \end{tabular}
    \caption{Duration of the full-ABM and hybrid model (10 runs) using the full population sample (100\%).}
    \label{table:duration_100pt}
\end{table} 

Now, knowing the least number of runs needed for both thresholds (2\% and 1\%), as well as the runtime for a single run, we can conclude that the total simulation time is smallest for the partial sample at the threshold 1\% and for the full population sample at the threshold 2\%.
The hybrid model requires at least $5.03$ hours of simulation time at the 2\% threshold (full population sample) and $15.28$ hours at the 1\% threshold (partial sample).
In comparison, the full-ABM requires $46.06$ hours at the 2\% threshold (full population sample) and $60.69$ hours at the 1\% threshold (partial sample).



\subsection{Extreme Cases} \label{subsec:extremeExperiments}
To assess the robustness of the hybrid coupling, we consider a set of extreme initial configurations. The purpose of these experiments is to examine whether the bidirectional exchange between the ABM and PDE components establishes epidemiological interaction across the interface and remains numerically stable under highly asymmetric conditions. People continue to follow their daily activities, and restrictions are imposed as before. The only change we make is to the initial health state of all individuals.
In the Susceptible-ABM scenario, all agents are initially susceptible, while the PDE population is initially exposed except for a single susceptible individual, and the exposed quickly become infectious. In the Exposed-ABM scenario, all agents start as exposed, whereas the PDE population is initially susceptible except for one exposed individual. This single individual ensures that the PDE population is initially distributed according to the landscape, and that subsequently entering individuals are added (or removed) based on the current distribution. Without this individual, new entrants would be uniformly distributed across the domain, which would not resolve quickly enough and could destabilize the model. An alternative would be to initialize the distribution of entering individuals directly based on the landscape.
In these experiments, we can observe the influence of commuting individuals from Berlin or Brandenburg on each other's infection dynamics. 

The agents' locations are plotted with their respective health states using different colors:
\textcolor{softgreen}{susceptible}, 
\textcolor{yellow}{exposed},
\textcolor{orange}{infectious}, 
\textcolor{red}{symptomatic}, 
\textcolor{darkred}{requires hospitalization}, 
\textcolor{purple}{critical}, 
\textcolor{gray}{recovered}.

\subsubsection{25\% Population}
\begin{figure}[h!]
    \centering
    \begin{subfigure}[b]{0.4\textwidth}
        \includegraphics[width=\linewidth]{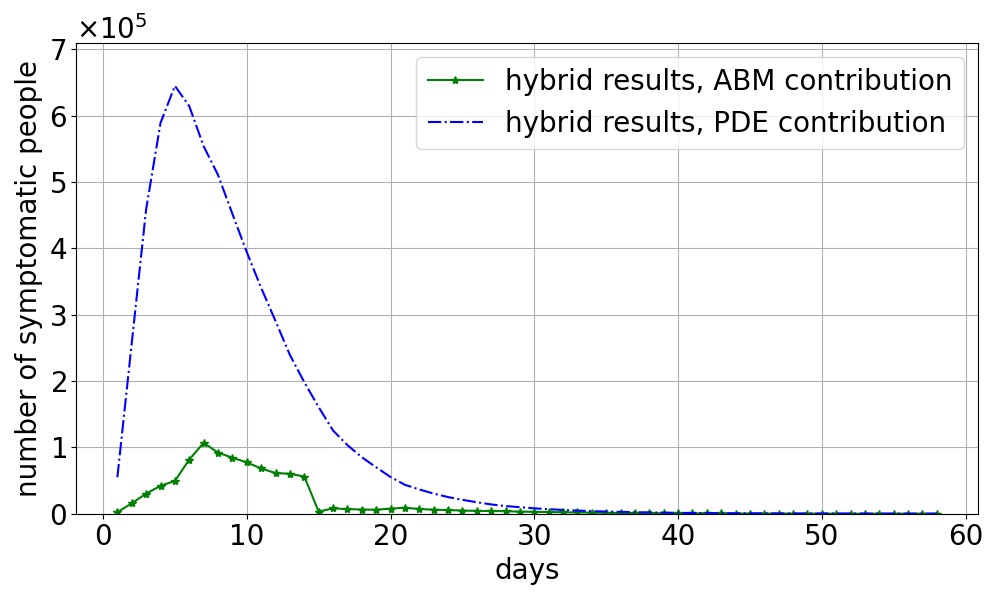}
        \includegraphics[width=\linewidth]{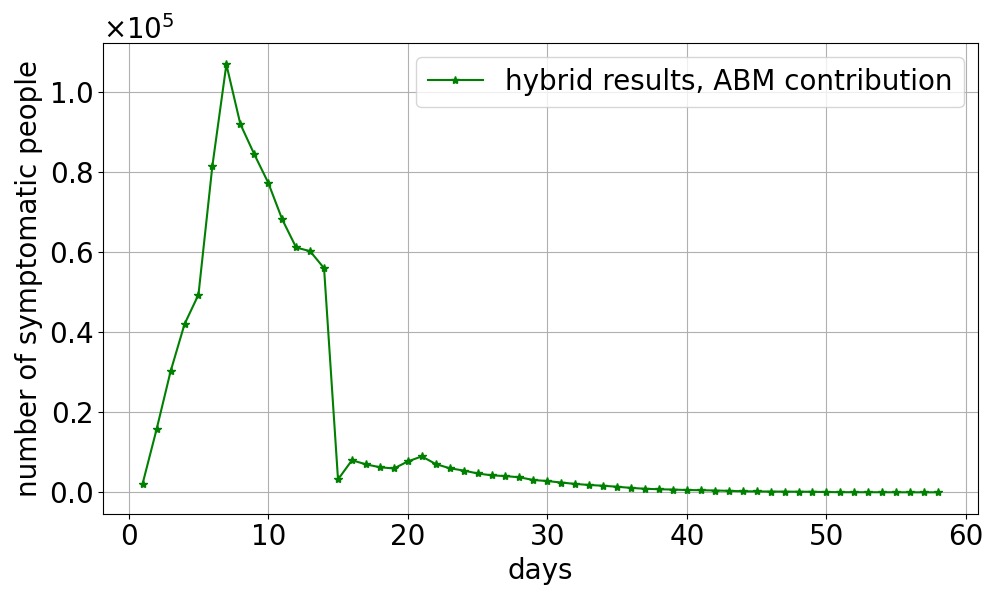}
        \caption{Total number of symptomatic people.}
        \label{fig:extreme_1_total_numbers_25pt}
    \end{subfigure}
    \hfill
    \begin{subfigure}[b]{0.58\textwidth}
        \includegraphics[width=\linewidth]{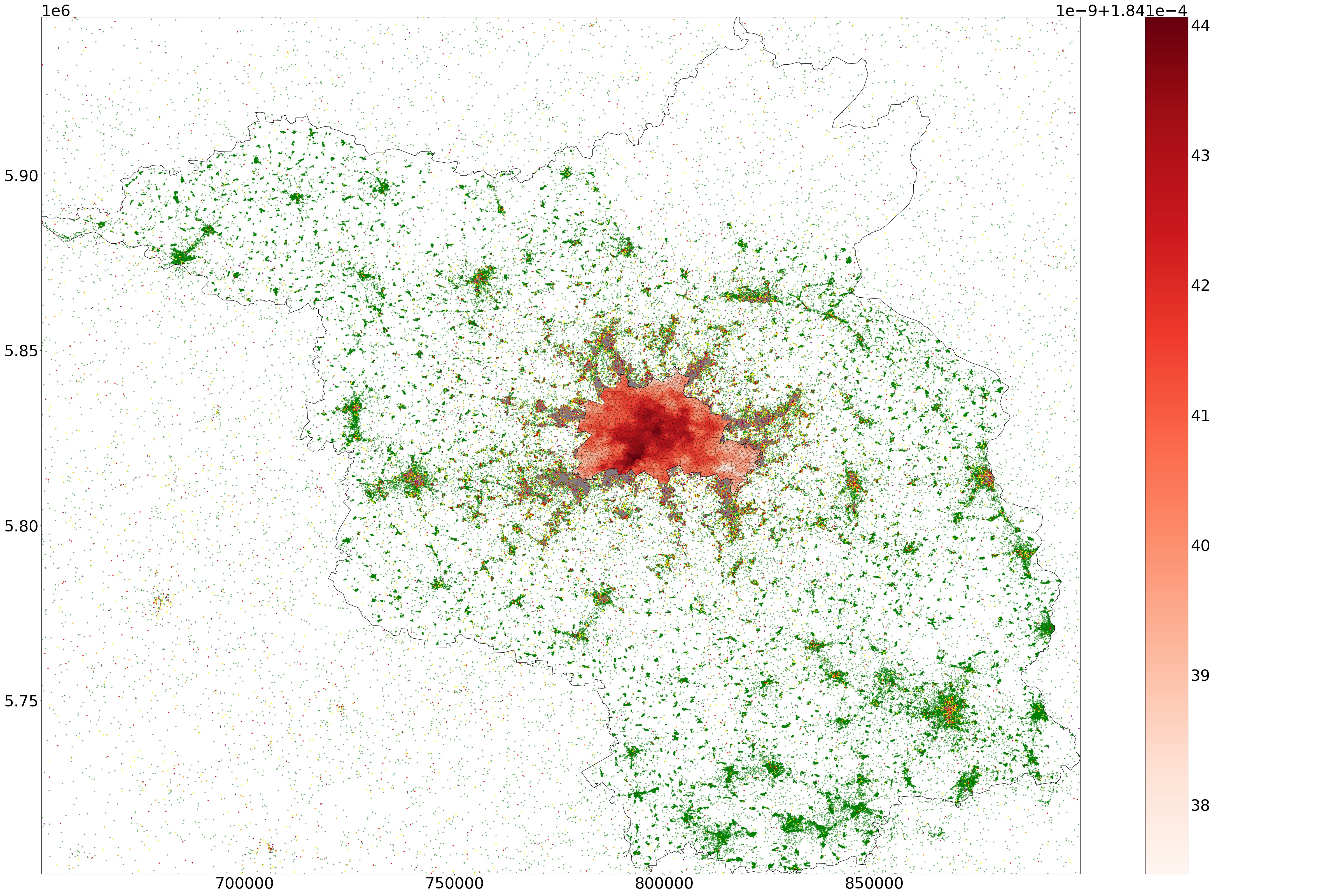} 
        \caption{Symptomatic density in Berlin and agents in Brandenburg on day 4 at 4:30 PM. The health state of agents is visualized with different colors.}
        \label{fig:extreme_1_density_agents_25pt}
    \end{subfigure}
    \caption{Everyone in PDE model contribution (\ref{eq:PDE-model}) of hybrid model is initially exposed and everyone in ABM contribution of hybrid model is initially susceptible. A 25\% population sample is used.}
\end{figure}

\begin{figure}[h!]
    \centering
    \begin{subfigure}{0.41\linewidth}
        \includegraphics[width=\linewidth]{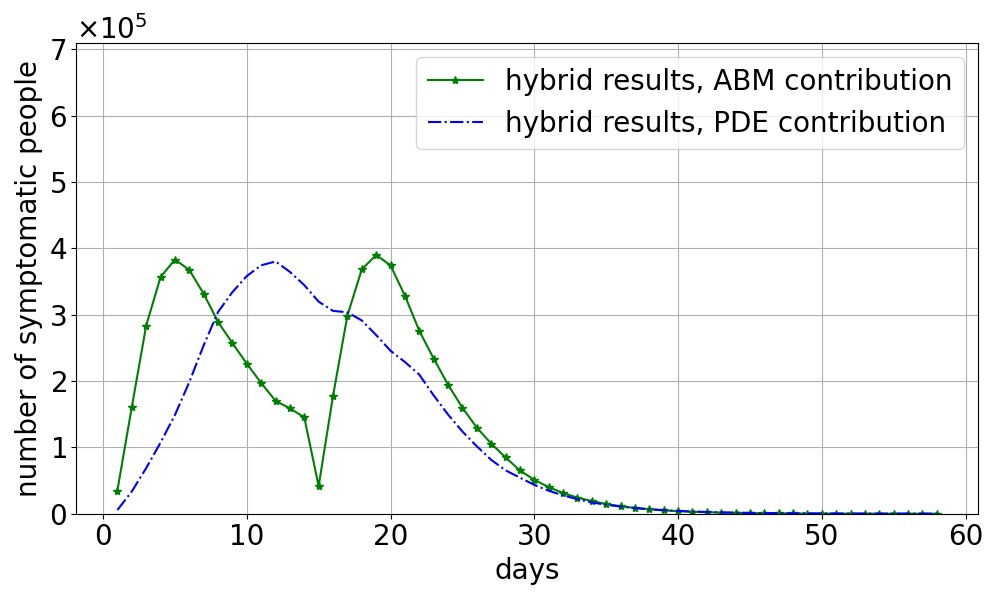}
        \caption{Total number of symptomatic people.}
        \label{fig:extreme_2_total_numbers_25pt}
    \end{subfigure}
    \begin{subfigure}{0.58\linewidth}
        \includegraphics[width=\linewidth]{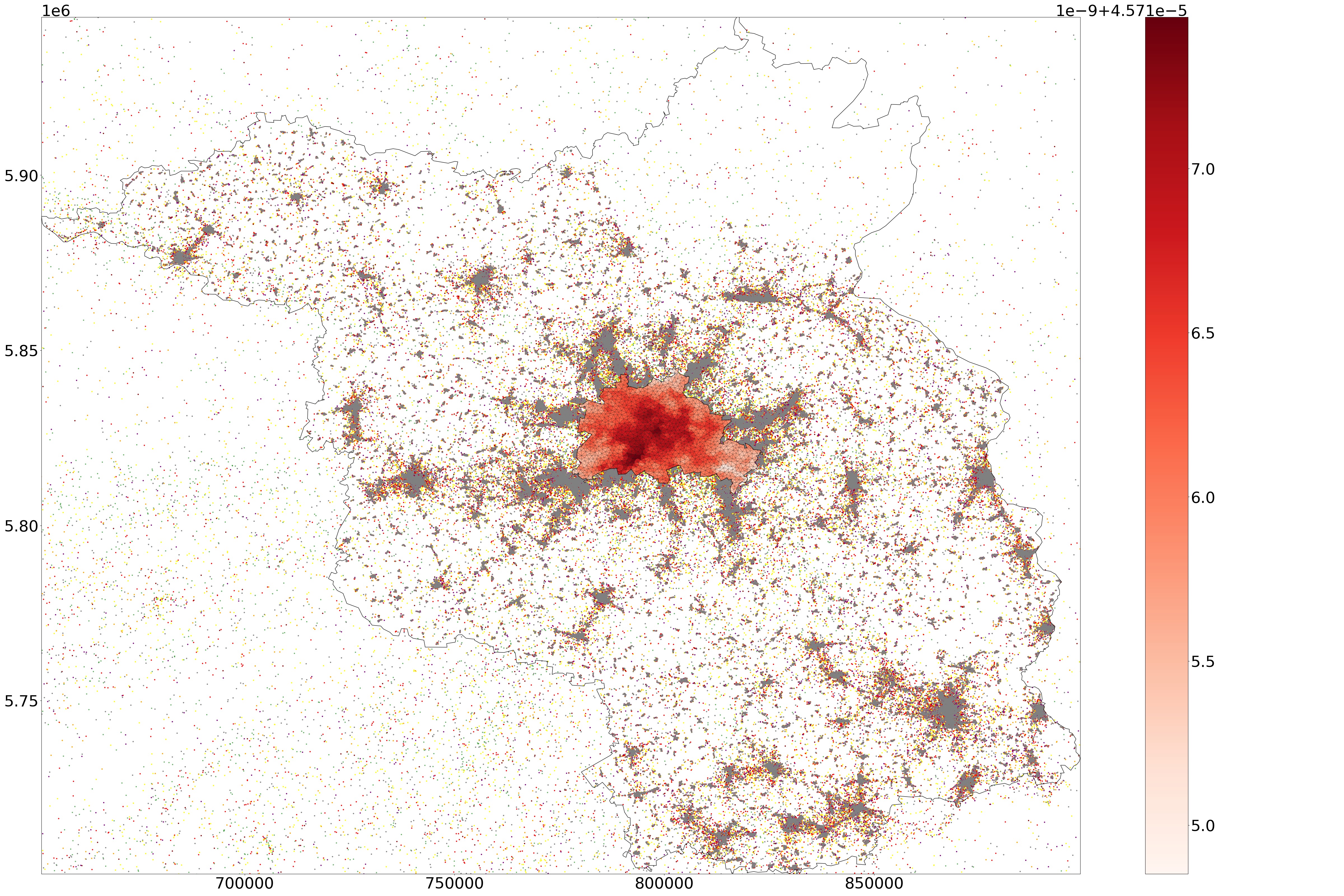}
        \caption{Symptomatic density in Berlin and agents in Brandenburg on day 4 at 8:30 PM. The health state of agents is visualized with different colors.}
        \label{fig:extreme_2_density_agents_25pt}
    \end{subfigure}
    \caption{Everyone in ABM contribution of hybrid model is initially exposed and everyone in PDE model contribution (\ref{eq:PDE-model}) of hybrid model is initially susceptible. A 25\% population sample is used.}
\end{figure}

The total number of symptomatic individuals for the ABM contribution and the PDE contribution (\ref{eq:PDE-model}) of the hybrid model are shown in Figs.~\ref{fig:extreme_1_total_numbers_25pt} and~\ref{fig:extreme_2_total_numbers_25pt}, while snapshots of the corresponding experiments can be found next to these plots in Figs.~\ref{fig:extreme_1_density_agents_25pt} and~\ref{fig:extreme_2_density_agents_25pt}. For better comparability, we set the maximum of the y-axis to the same value in the total number plots, except for the close-up.

Examining the total number of symptomatic individuals in these plots, we observe a significant difference in how the states influence each other. 
In the Susceptible-ABM scenario (see Fig.~\ref{fig:extreme_1_total_numbers_25pt}), the peak number of symptomatic individuals in Berlin (blue) is significantly higher than in Brandenburg, to the extent that the small wave in Brandenburg is barely comparable. 
In the Exposed-ABM scenario (see Fig.~\ref{fig:extreme_2_total_numbers_25pt}), the peaks have shifted toward each other: the peak in Berlin is lower than in the Susceptible-ABM scenario, while the peak in Brandenburg is now clearly higher. 
However, the peak in Brandenburg remains significantly lower than the previous peak in Berlin from the Susceptible-ABM scenario. It is evident that there is a notable difference depending on which model the entire population except one individual is set to exposed or susceptible.

Additionally, we observe the formation of a second wave in Brandenburg in the Exposed-ABM scenario, with a sharp decline followed by a significant increase in symptomatic cases between the waves. However, upon closer inspection of the data (see Fig.~\ref{fig:extreme_1_total_numbers_25pt}), this pattern is also present in the Susceptible-ABM scenario in the ABM, where a small second wave, if we may call it that, forms in a similar manner. The increases and decreases in the Susceptible-ABM scenario in the ABM, where we observe very few cases, seem to be partially driven by the activity data (see Fig.~\ref{fig:activityChanges}). However, this is less the case for the other scenarios, where the waves are more pronounced, and the number of symptomatic cases is significantly higher, or the first wave emerges later. Nevertheless, it is possible that the activity data triggered the formation of the second wave. Interestingly, in the Susceptible-ABM scenario for the ABM, the number of symptomatic cases even reaches almost 0.0 between the waves.

\begin{figure}[h!]
    \centering
    \includegraphics[width=0.5\linewidth]{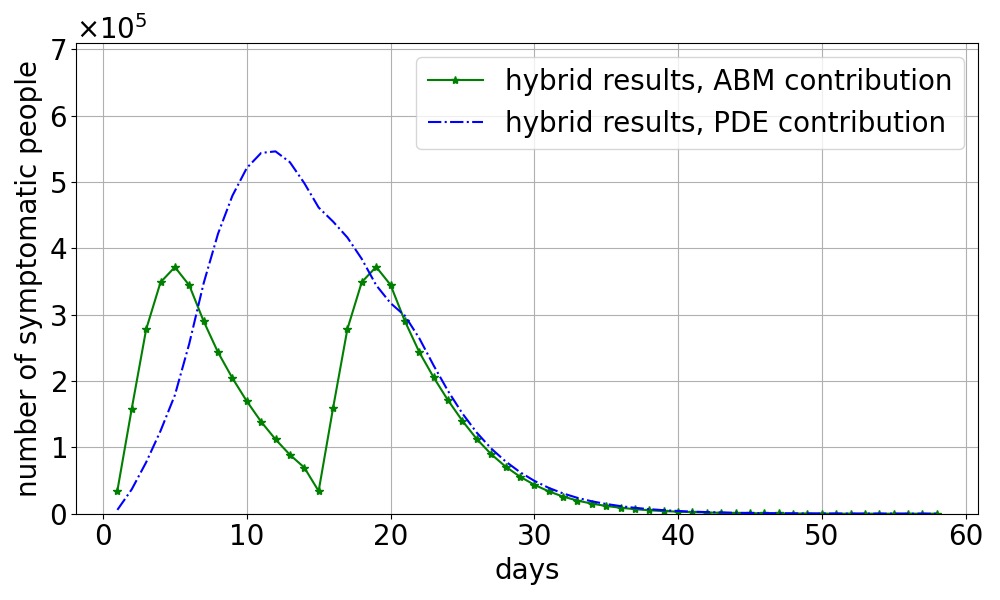}
    \caption{Total number of symptomatic people. Everyone in ABM contribution of hybrid model is initially exposed and everyone in PDE model contribution (\ref{eq:PDE-model}) of hybrid model is initially susceptible. The transition of agents from the PDE domain to the ABM domain has been removed.}
    \label{fig:extreme_2_total_numbers_25pt_less_coupling}
\end{figure}

The peak in Berlin occurs on day 12 (see Fig.~\ref{fig:extreme_2_total_numbers_25pt}), shortly before the second time interval with the lower infection rate begins. It is worth noting that the activity change rates, together with the increasing cases in Berlin, may have contributed to the formation of a second wave. After adjusting the implementation to prevent individuals from jumping out of the PDE domain into the ABM (see Fig.~\ref{fig:extreme_2_total_numbers_25pt_less_coupling}), a second wave is still visible, and it appears almost identical in the case of the ABM compared to the previous experiment with full coupling. Hence, the influence of Berlin is small or not clearly visible in this case. Additionally, as the individuals in the PDE domain are required to stay in Berlin, symptomatic cases have increased in Berlin.

%
\subsubsection{100\% Population}
\begin{figure}[h!]
    \centering
    \begin{subfigure}[b]{0.4\textwidth}
        \includegraphics[width=\linewidth]{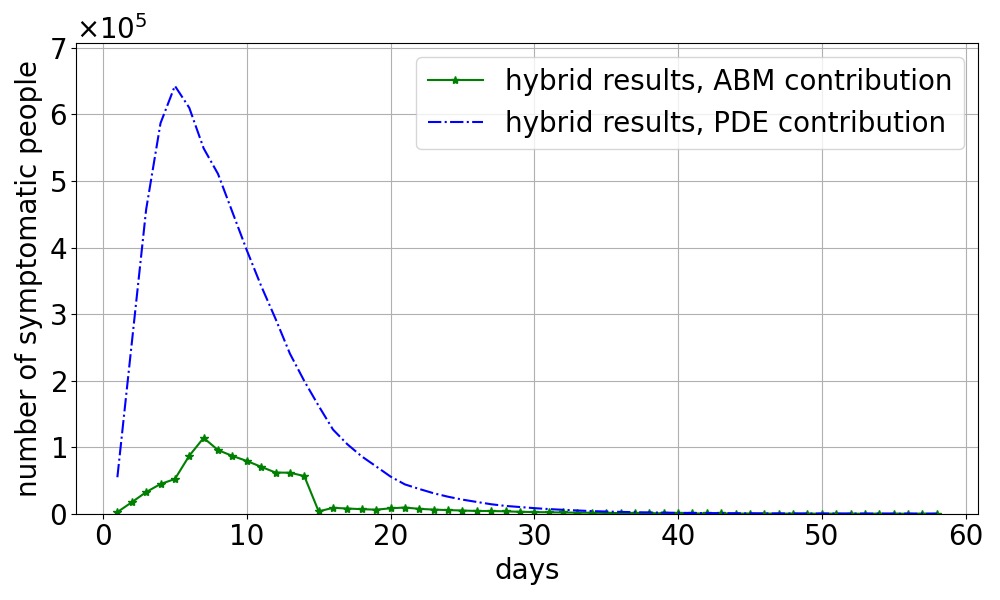}
        \includegraphics[width=\linewidth]{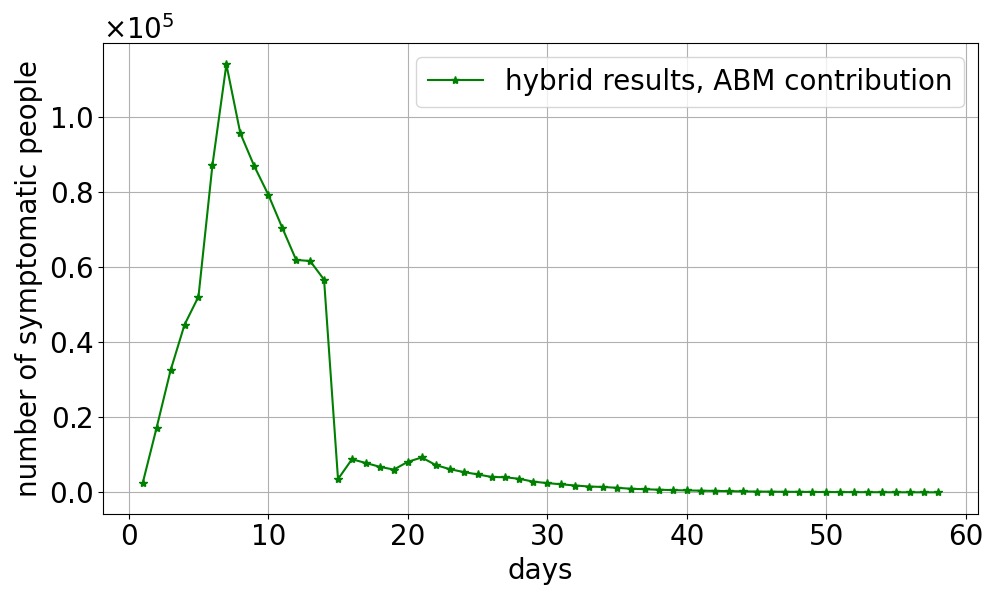}
        \caption{Total number of symptomatic people.}
        \label{fig:extreme_1_total_numbers_100pt}
    \end{subfigure}
    \hfill
    \begin{subfigure}[b]{0.58\textwidth}
        \includegraphics[width=\linewidth]{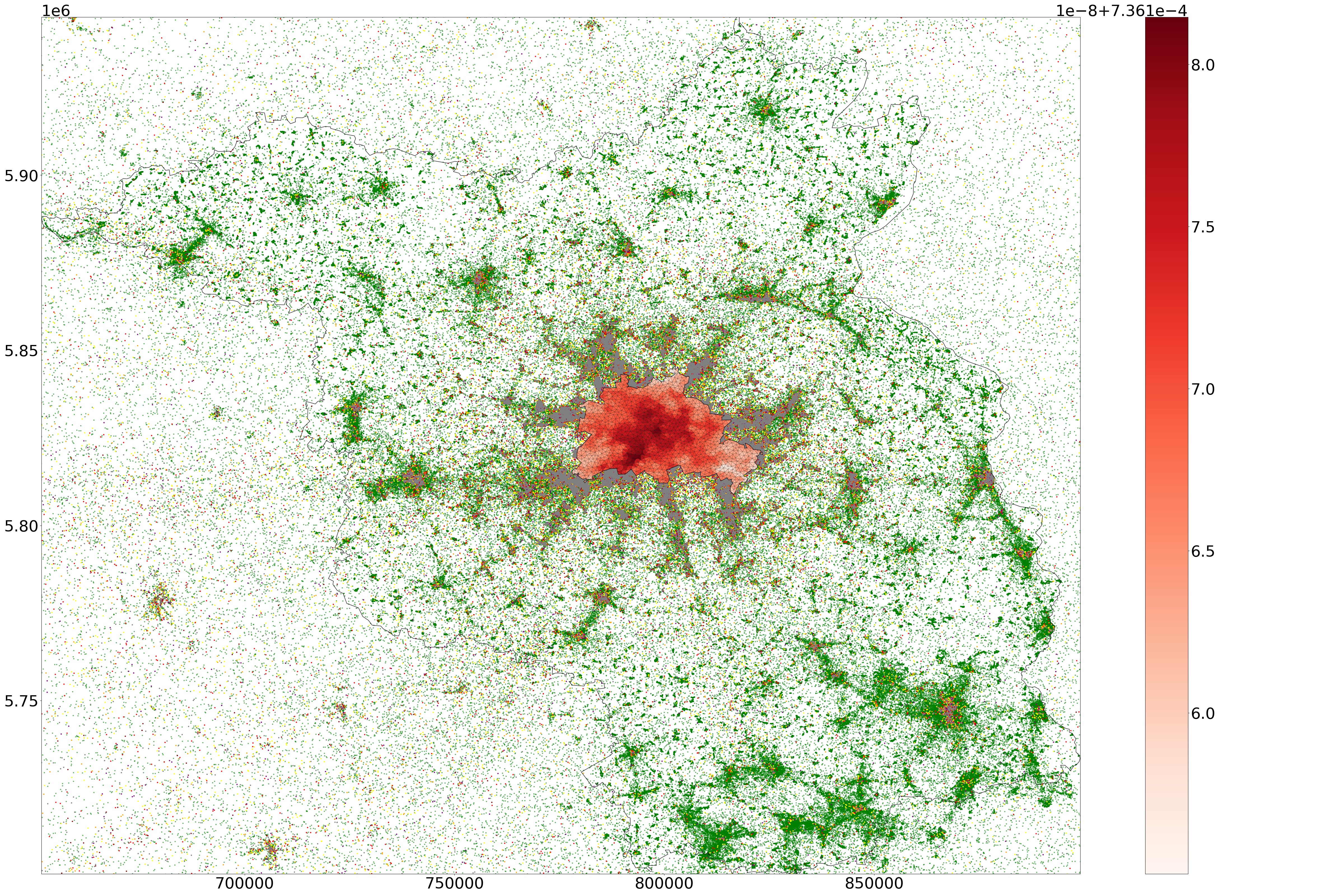} 
        \caption{Symptomatic density in Berlin and agents in Brandenburg on day 4 at 4:30 PM. The health state of agents is visualized with different colors.}
        \label{fig:extreme_1_density_agents_100pt}
    \end{subfigure}
    \caption{Everyone in PDE model contribution (\ref{eq:PDE-model}) of hybrid model is initially exposed and everyone in ABM contribution of hybrid model is initially susceptible. The full population sample (100\%) is used.}
\end{figure}

\begin{figure}[h!]
    \centering
    \begin{subfigure}{0.41\linewidth}
        \includegraphics[width=\linewidth]{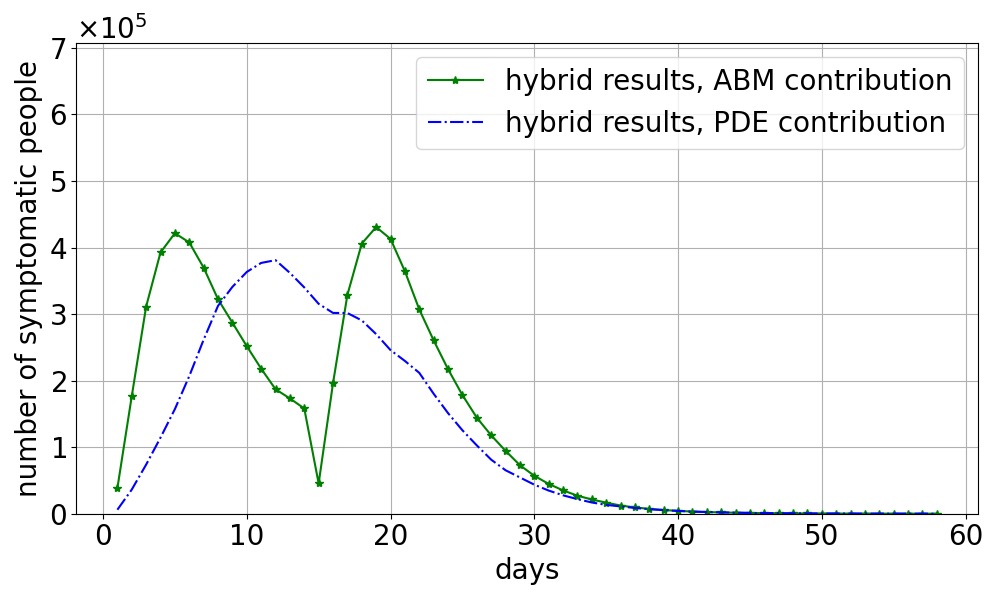}
        \caption{Total number of symptomatic people.}
        \label{fig:extreme_2_total_numbers_100pt}
    \end{subfigure}
    \begin{subfigure}{0.58\linewidth}
        \includegraphics[width=\linewidth]{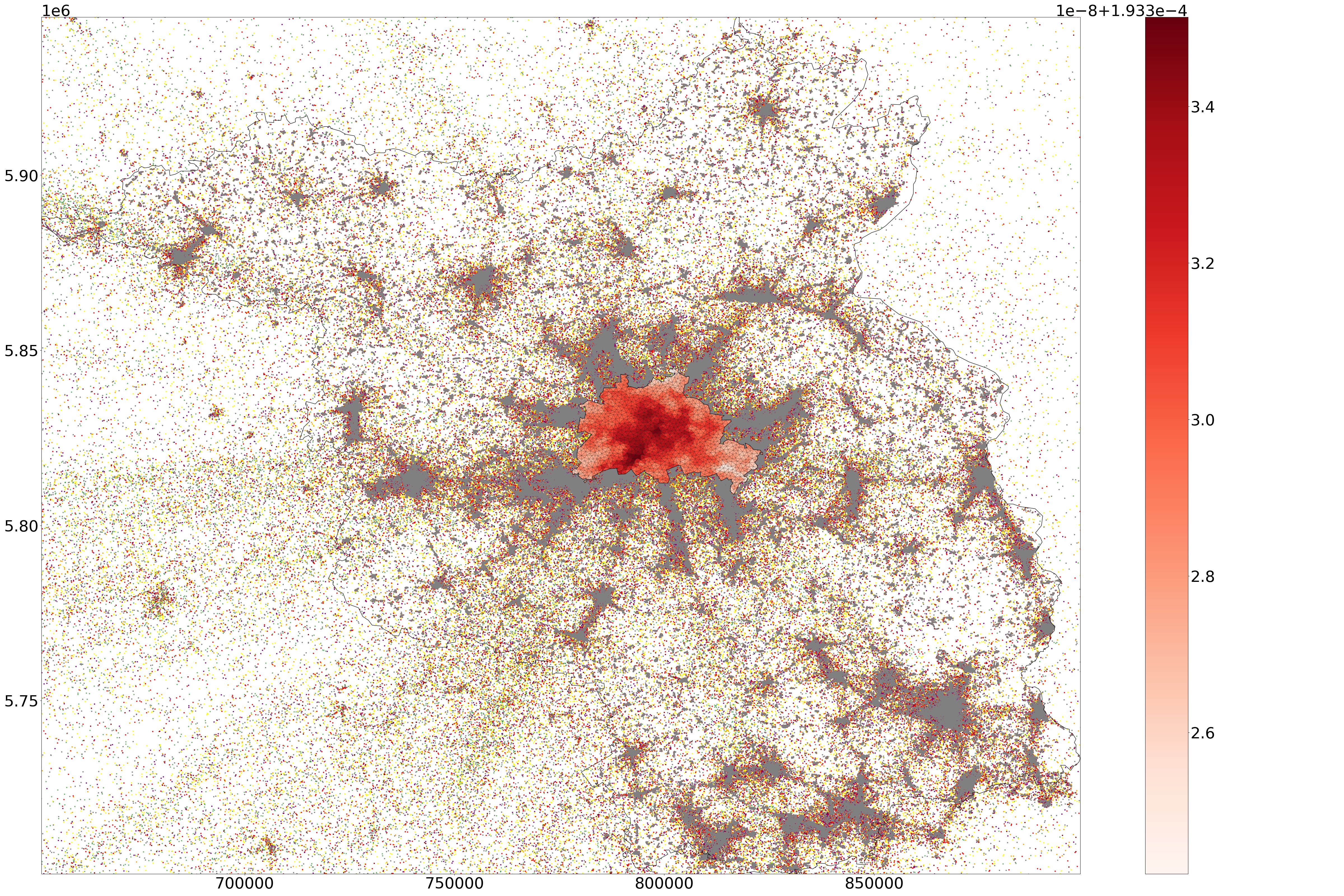}
        \caption{Symptomatic density in Berlin and agents in Brandenburg on day 4 at 8:30 PM. The health state of agents is visualized with different colors.}
        \label{fig:extreme_2_density_agents_100pt}
    \end{subfigure}
    \caption{Everyone in ABM contribution of hybrid model is initially exposed and everyone in PDE model contribution (\ref{eq:PDE-model}) of hybrid model is initially susceptible. The full population sample (100\%) is used.}
\end{figure}

Similarly, the total number of symptomatic individuals for the ABM and PDE contributions (\ref{eq:PDE-model}) of the hybrid model are presented in Figs.~\ref{fig:extreme_1_total_numbers_100pt} and~\ref{fig:extreme_2_total_numbers_100pt}. Corresponding experiment snapshots are displayed next to these plots in Figs.~\ref{fig:extreme_1_density_agents_100pt} and~\ref{fig:extreme_2_density_agents_100pt}. Again, to ensure comparability, the y-axis maximum has been set to the same value across the total number plots, except for the close-up.

The upper and lower plots in Fig.~\ref{fig:extreme_1_total_numbers_100pt} appear identical to the upper and lower plots in Fig.~\ref{fig:extreme_1_total_numbers_25pt}, respectively. 
The peak of the ABM curve is higher for the full population, but it is important to note that this result is based on a single run. Once again, the increases and decreases, particularly in the Susceptible-ABM scenario within the ABM, appear to be partially driven by the activity data (see Fig.~\ref{fig:activityChanges}).

Looking at the total numbers plot of the Exposed-ABM scenario (see Fig.~\ref{fig:extreme_2_total_numbers_100pt}), it appears similar to the plot for the 25\% population sample (see Fig.~\ref{fig:extreme_2_total_numbers_25pt}). Indeed, both plots are quite similar, with a slightly higher overall number of symptomatic individuals. Notably, both peaks are somewhat higher but occur on the same days (5 and 19).

\section{Discussion}
Our hybrid modeling approach, which integrates an agent-based model (ABM) with a partial differential equation (PDE) model, offers several advantages but also introduces new challenges. A key limitation lies in the deterministic nature of the PDE component, meaning that, e.g., variability in disease duration and the effect of fluctuating contact opportunities on transmission dynamics are not present in this part of the model. On the other hand, the hybrid coupling introduces new sources of variability through information loss, such as a recovered agent entering Berlin and later reappearing as a susceptible agent, which introduces changes to the theoretical dynamics (see Fig.~\ref{fig:compartments_flow}).
As we scale the 25\% population results to the full population, the variance will scale accordingly, resulting in a larger variance for the scaled 25\% population results.
Additionally, changing the population size affects the maximum number of agents present in a facility of a given category. For the full population, this number is higher for the same facility compared to the partial sample. As a result, the probability of infection decreases when the same number of people are inside the facility. This lower probability of infection may contribute to a lower overall variance. 
When considering the total population, the probability of infection fluctuates less when an infectious agent leaves a facility compared to the partial population sample, making the changes less drastic. At the same time, incorporating more people into the model increases mixing. Given the complexity of these effects, the most reliable approach was to conduct simulations.

The deterministic PDE dynamics and the coupling process both reduce variability in the 25\% and the 100\% scenarios. 
In fact, the hybrid model introduces a different type of noise due to the abstraction and simplification of agent-level dynamics. Agent identifiers are lost in the PDE region, leading to mismatches when assigning states upon re-entry to the ABM. For instance, an agent may reappear earlier or later than expected, or with an incorrect health status. To mitigate this, we propose future improvements such as storing metadata when an agent transitions into the PDE domain -- including expected return time and current health status. Upon re-entry, this information could be used to assign a health state based on time spent in the PDE region and the transition dynamics from the compartment model (see Fig.~\ref{fig:compartments_flow}). This strategy would help reduce artificial variance introduced by random reassignment. Additionally, stochasticity could be reintroduced into the PDE model itself -- for example, through noise terms in the compartment equations -- to better reflect variability in transitions.

We used a 25\% sample of the full dataset, which is also employed for calibration and evaluation in existing studies~\cite{Muller2020realistic,Muller2021ABM}. For readers interested in population sampling strategies, we refer to the general context of population sampling~\cite{AHMED2024sampling, Ali2024Sampling} and in the context of mobility ABMs~\cite{Bigi2024Synthetic, Caglar2022Synthetic}. 
In scenarios where detailed mobility data such as mobile phone trajectories are not available, alternative approaches may be employed. For example, radiation-type models~\cite{Massimiliano2022Modeling} can estimate commuter flows and can be integrated into epidemic frameworks. However, due to their parameter-free structure, such models may have limited flexibility in capturing heterogeneous mobility patterns. 

Interestingly, while the number of cases in Berlin has little influence on the infection wave in Brandenburg, there is a strong influence in the other direction. This asymmetry is especially visible in our plots when varying the population size, for example, in Figs.~\ref{fig:extreme_1_total_numbers_25pt} and~\ref{fig:extreme_1_total_numbers_100pt}, but especially through the extreme experiments within the same population size. 
It is important to keep in mind that the population size in Berlin is significantly larger than in Brandenburg and also larger than the total number of agents in the ABM contribution of the hybrid model.
One explanation lies in the asymmetry of commuting: according to data from 2021, around 270,000 inhabitants of Brandenburg had their work place in Berlin, while only around 108,000 inhabitants of Berlin had their work place in Brandenburg~\cite{Pendlerzahlen2021_2022Pendleratlas}. These numbers likely underestimate the actual pre-pandemic commuting flows due to increased remote work in 2021. Still, they highlight the significant directional influence of Brandenburg on Berlin. 

A comparison with a purely PDE-based formulation was considered. Modeling both domains within a single spatially continuous PDE framework, or alternatively introducing two coupled PDE systems, are possible. However, Brandenburg exhibits a sparse population distribution, and commuter movements involve daily returns to the home federal state, where infection transmission may occur. Capturing such effects would require a more complex PDE formulation. In particular, chemotaxis-type PDE infection models~\cite{Colli2024Chemotaxis} with drift terms could reproduce asymmetric commuter influences. As our primary objective is to assess the hybrid ABM–PDE coupling while preserving explicit trajectory information in the heterogeneous region, we did not include a full-PDE comparison.

From a technical perspective, there are several opportunities for performance improvements, e.g., for better runtime and better memory efficiency. For instance, the ABM was not implemented in the most efficient way but rather in the same manner as the ABM component of the hybrid model. The hybrid model itself also offers room for optimization, such as interrupting the ABM event data evaluation to perform a PDE step and exchange information between both models. This would allow the use of the entire event dataset instead of creating and working with sliced event data, as done in this paper.
Rather than creating artificial event slices, a more efficient approach could be to store the index of the last evaluated event for each agent and compute their position inside a facility at the beginning of each time step during runtime. Additionally, by pre-labeling facilities within the PDE region, it would be possible to determine more quickly whether an agent is entering the PDE region. If an agent is commuting, a lookup could be performed to check whether the transition is from a facility to the PDE region, allowing for a faster decision on whether a jumping process should occur. The precise implementation details still need to be determined.


\section{Conclusion}
In this paper, we presented a novel hybrid modeling approach that couples agent-based models (ABMs) with partial differential equations (PDEs) to efficiently simulate epidemiological dynamics across heterogeneous geographic regions. Our approach leverages the individual-level detail of ABMs while incorporating the spatially continuous representation of PDEs, allowing for a more computationally efficient yet spatially resolved simulation framework. Our results demonstrate how the integration of these two components improves model efficiency and highlights its ability to capture key epidemiological dynamics, such as commuting-driven transmission between Berlin and Brandenburg.

By applying this hybrid ABM-PDE model to the Berlin-Brandenburg region using real-world mobility and infection data, we demonstrated its ability to capture the distinct epidemiological dynamics of urban and rural areas. The model successfully reduced computational cost significantly compared to a full-ABM simulation. This efficiency gain becomes evident when comparing the simulation times and error margins of the hybrid model and the full-ABM. 
Specifically, the hybrid model achieved shorter simulation times and lower errors than the ABM, regardless of whether a partial (25\%) or full (100\%) population sample was used. Thus, our hybrid approach is a practical alternative for real-time epidemiological assessments. A key advantage over full-ABM models such as EpiSim is that our approach allows for efficient simulations of 100\% of the population on a standard computer, making large-scale epidemiological modeling more accessible and computationally feasible.

Looking ahead, we aim to extend this hybrid framework into a fully integrated ABM-PDE-ODE system, allowing for even more flexible and modular modeling of complex disease dynamics across varying spatial and structural scales.

\section*{Acknowledgments} 
We acknowledge Kai Nagel for valuable discussions on the agent-based model. We also thank Jakob Benjamin Rehmann and Sydney Cornelia Paltra for providing essential data for the Berlin and Brandenburg experiments and for assistance with their software. 
We would like to thank Inan Bostanci for the many meetings and constructive feedback throughout the process.

This work was funded by the German Federal Ministry of Research, Technology and Space (BMFTR) through the project EPISERVE (project ID: 031L0324A) and through the project MODAL (fund numbers 05M2025). EPISERVE is part of the German Modeling network for severe infectious diseases (MONID). Further funding was provided by the Deutsche Forschungsgemeinschaft (DFG, German Research Foundation) under Germany´s Excellence Strategy – The Berlin Mathematics Research Center MATH+ (EXC-2046/1, EXC-2046/2, project ID: 390685689). 

\clearpage

\bibliography{references}{}
\bibliographystyle{plain}



\newpage
\appendix

{\Large Appendices}

\section{Berlin – PDE-Based Compartment Model} \label{appendix:PDEModelDerivation}

We proceed as follows: we first define a motion-based ABM with a single health state using a landscape to guide spatial movement. We then extend the model to include multiple health states and associated transitions between them. Next, we translate (reduce) this into a PDE-based compartment model, and finally, we present its weak formulation to enable numerical simulation using the finite element method.

\subsubsection{Construction of the Motion-Based ABM} \label{subsubsec:motionABM}
The overall aim is to construct an ABM consisting of agents moving stochastically and continuously within a domain, described by Brownian motion and a landscape in which agents can change their health status probabilistically, influenced by the health status of their neighbors. The first step is to construct a motion-based ABM in which agents move stochastically and continuously within a spatial domain. Each agent's movement is governed by Brownian motion influenced by a landscape potential, which guides them toward frequently visited areas. So for now, we focus only on mobility - no health state transitions are considered yet. The domain $\Omega_{Be} \subseteq \mathbb{R}^2$ represents the urban area of Berlin. The landscape $V$ defines spatial preferences and is specified on an equidistant grid. Because agents move in continuous space, the local value of $V$ at their current grid cell influences the agent's next movement.

To model the agent's movement, we consider an overdamped Langevin diffusion process~\cite{Li2023Numerical} following the approach of Helfmann et al.~\cite{Helfmann2021Interacting}, which describes the evolution of an agent's position $X(t) \in \mathbb{R}^2$ over time $t$ as:
\begin{align*}
    dX(t) = -\nabla V(X(t)) dt + \sqrt{2D} \, dB(t),
\end{align*} 
where $V: \Omega_{Be} \rightarrow \mathbb{R}$ is the landscape potential, $D \in \mathbb{R}$ is the diffusion coefficient, and $B(t) \in \mathbb{R}^2$ denotes a standard Brownian motion. The drift term $-\nabla V(X(t))$ causes agents to preferentially move toward regions of lower potential, which correspond to areas with higher empirical presence in the underlying data.

We exploit the property of the system that the landscape $V: \; \Omega_{Be} \rightarrow \mathbb{R}$ can be expressed as
\begin{align} \label{eq:landscape_definition}
    V(X) = - \biggl(\frac{D}{2}\biggr) \log \bigl( p_{st}(X) \bigr),
\end{align}
where $p_{st}(X)$ is the 
probability distribution of the agents at location $X$~\cite{Li2023Numerical}.
To approximate the probability distribution of the agents at each location, we use mobile phone data. 
First, we generate agent trajectories by determining their location for each hour of the day. 
Next, we aggregate all trajectories for each hour of the day over the entire week to construct a histogram. After normalization, this histogram represents the probability distribution of the agents, which we use to compute the landscape. Mobile phone data is available for Berlin and Brandenburg, so the landscape is initially defined for the entire region, and then we extract the portion corresponding to Berlin.

Our motion-based ABM for agent $i=1,2,\dots,N$ is then given by the overdamped Langevin diffusion equation system
\begin{align*}
    dX_i(t) = -\nabla V(X_i(t)) dt + \sqrt{2D} \, dB_i(t).
\end{align*}

\subsubsection{Extension with Health States} \label{subsubsec:statusABM}
We now extend the motion-based ABM by introducing multiple health states and transition rules to capture the dynamics of infectious disease progression. The population is divided into subpopulations, each associated with a distinct epidemiological status. These statuses and their transitions are modeled via a set of interaction rules $\mathcal{R} = (\mathcal{R}_r)_{r=1}^{10}$, following Helfmann et al.~\cite{Helfmann2021Interacting} and the work by Müller and colleagues~\cite{Muller2020realistic,Muller2021ABM} for the epidemiological assumptions:

\begin{equation}\label{ABMrules}
\begin{aligned}
  & \mathcal{R}_1: &\quad T_S + T_{E \cup I} &\rightarrow 2 T_E  &&&&&&&&&&&&&& \mathcal{R}_6: &\quad T_{\sY} &\rightarrow T_H \\
  & \mathcal{R}_2: &\quad T_E &\rightarrow T_I &&&&&&&&&&&&&& \mathcal{R}_7: &\quad T_H &\rightarrow T_R \\
  & \mathcal{R}_3: &\quad T_I &\rightarrow T_R &&&&&&&&&&&&&& \mathcal{R}_8: &\quad T_H &\rightarrow T_C \\
  & \mathcal{R}_4: &\quad T_I &\rightarrow T_{\sY} &&&&&&&&&&&&&& \mathcal{R}_9: &\quad T_C &\rightarrow T_{H_C} \\
  & \mathcal{R}_5: &\quad T_{\sY} &\rightarrow T_R &&&&&&&&&&&&&& \mathcal{R}_{10}: &\quad T_{H_C} &\rightarrow T_R,
\end{aligned}
\end{equation}
where $T_Y$ describe the different agent types with $Y\in \{S,E,I,\sY,H,C,H_C,R\}$. The health statuses are defined as follows:
\begin{itemize}
    \item susceptible $\left(S\right)$
    \item exposed $\left(E\right)$ - not symptomatic, not infectious
    \item infectious $\left(I\right)$ - not symptomatic, infectious
    \item symptomatic $\left(\,\sY\right)$ - symptomatic, infectious
    \item requiring hospitalization (after being symptomatic) $\left(H\right)$ 
    \item critical $\left(C\right)$ 
    \item requiring hospitalization (after being in critical state) $\left(H_C\right)$
    \item recovered $\left(R\right)$.
\end{itemize}
The agent type or state of agent $i$ at time $t$ is defined as $Y_i(t)$. 
The total number of people requiring hospitalization is the sum of agents that are in state $H$ or $H_C$. 
The total number of infected people is the sum of agents that are in state $E$, $I$, $\sY$, $H$, $C$ or $H_C$. 
We define the total number of all agents as $N$. 
We do not consider deaths or births, i.e., our total population $N$ remains constant with respect to time.

State transitions (health status change) occur at discrete time points $t_k$, $k=1,2,\dots,K$, with constant time step size $\Delta t$. 
The probability of a health status change for agent $i$ at time $t_k$ is given by:
\begin{align*}
    &&1-\exp{ 
    \left(
    -\Delta t \sum_{r \in \mathcal{R}} \lambda_i^r(t_k) 
    \right)
    }, && k=1,\dots,K,
\end{align*}
where the 
transition rate function $\lambda_i^r(t)$ of agent $i=1,\dots,N$ and rule $r=1,\dots,10$ at time $t$ 
is given by
\begin{align*}
    \lambda_i^r(t) = \mathcal{T}_r \mathbbm{1}_{Y'} \bigl(Y_i(t)\bigr) \sum_{j=1}^N A_{ij}(t) 
    \mathbbm{1}_{Y''} \bigl(Y_j(t)\bigr), && Y',Y'' \in \{S,E,I,\sY,H,C,H_C,R\}
\end{align*} 
\cite{Helfmann2021Interacting}.
Further, we have the constant transition rates $\mathcal{T}_r$ for each rule $r$, characteristic function $\mathbbm{1}_{\{ \cdot \}}$, and an adjacency matrix $A(t) = \left( A_{ij}(t) \right)_{i,i=1}^{N}$ with 
\begin{align*}
    A_{ij}(t) = \left\{ \begin{array}{ll}
1 & \text{if } ||X_i(t)-X_j(t)||\leq d_{\text{int}} \text{ for } i \neq j  \\
0 & \text{else}
\end{array}
\right.
\end{align*}
where $d_{\text{int}}$ is a predefined distance threshold, such that agents are considered close enough to transmit the disease. 
If multiple transitions are possible for an agent, the specific rule is chosen with probability 
\begin{align*}
    \frac{\lambda_i^r(t_k)}{\sum_{r \in \mathcal{R}} \lambda_i^r(t_k)}.
\end{align*}

In some cases, such as for \textit{susceptible} ($S$) or \textit{critical} ($C$) agents, only one rule applies, so the next state is uniquely determined (to \textit{exposed} ($E$) or \textit{requiring hospitalization} ($H_C$), respectively).

\subsubsection{Derivation of the PDE System} 

Each compartment in the PDE model represents the spatial density of individuals in a given health state, evolving over time. Health state transitions are incorporated through reaction terms in the PDEs, analogous to compartmental models. For example, the transition from exposed $E$ to infectious $I$ is governed by a loss term $\sigma E(x,t)$ in the equation for $E$, and a corresponding gain term in the equation for $I$. Similarly, new infections appear as a loss in the susceptible density and a gain in the exposed compartment, driven by contact with nearby infectious individuals. This structure allows the PDE model to capture both spatial movement and disease progression through the population. A schematic overview of the compartment structure and health state transitions is shown in Figure~\ref{fig:compartments_flow}.

Building on the rules (\ref{ABMrules}) from Appendix~\ref{subsubsec:statusABM} and following the approach by Helfmann et al.~\cite{Helfmann2021Interacting}, we now derive the system of stochastic PDEs governing the spatially distributed health compartments. The resulting stochastic PDE system reflects the combined effects of agent motion (via diffusion and drift in the landscape), state transitions (as reaction terms), and intrinsic stochasticity (representing random fluctuations in agent movement and interactions, modeled through multiplicative white noise terms in the PDE system).



Following this structure, the full system of equations is given as:
\begin{flalign*} 
    \frac{\partial S(x,t)}{\partial t} = \;& D \Delta S(x,t) + \divergence \left(\nabla V(x) S(x,t)\right) + \sqrt{2D} \nabla \left(\sqrt{S(x,t)} Z^D(x,t) \right) \\ & \hspace{-1cm} - \beta S(x,t) \left( I(t)^{B_r(x)} + \sY(t)^{B_r(x)} \right) 
    \\ & \hspace{-1cm} - \sqrt{\beta S(x,t)\left( I(t)^{B_r(x)} + \sY(t)^{B_r(x)} \right)} Z_1^I(x,t) \\
    \frac{\partial E(x,t)}{\partial t} = \; &D \Delta E(x,t) + \divergence \left(\nabla V(x) E(x,t)\right) + \sqrt{2D} \nabla \left(\sqrt{E(x,t)} Z^D(x,t) \right) \\ & \hspace{-1cm}+ \beta S(x,t) \left( I(t)^{B_r(x)} + \sY(t)^{B_r(x)} \right) 
    \\ & \hspace{-1cm} + \sqrt{\beta S(x,t)\left( I(t)^{B_r(x)} + \sY(t)^{B_r(x)} \right)} Z_1^I(x,t) \\ 
    & \hspace{-1cm}- \sigma E(x,t) - \sqrt{\sigma E(x,t)} Z_2^I(x,t) \\
   \frac{\partial I(x,t)}{\partial t} = \; &D \Delta I(x,t) + \divergence \left(\nabla V(x) I(x,t)\right) + \sqrt{2D} \nabla \left(\sqrt{I(x,t)} Z^D(x,t) \right)  \\ & \hspace{-1cm} + \sigma E(x,t) + \sqrt{\sigma E(x,t)} Z_2^I(x,t) - \phi_i I(x,t) 
   - \sqrt{\phi_i I(x,t)} Z_3^I(x,t) \\ & \hspace{-1cm} - \gamma I(x,t) - \sqrt{\gamma I(x,t)} Z_4^I(x,t) \\
   \frac{\partial \sY(x,t)}{\partial t}= \; &D \Delta \sY(x,t) + \divergence \left(\nabla V(x) \sY(x,t)\right) \\
   &\hspace{-1cm} + \sqrt{2D} \nabla \left(\sqrt{\sY(x,t)} Z^D(x,t) \right) + \gamma I(x,t) + \sqrt{\gamma I(x,t)} Z_4^I(x,t) \\ & \hspace{-1cm}- \phi_{sy} \sY(x,t) - \sqrt{\phi_{sy} \sY(x,t)} Z_5^I(x,t) - \eta \sY(x,t) - \sqrt{\eta \sY(x,t)} Z_6^I(x,t) \\
   \frac{\partial H(x,t)}{\partial t} = \; &D \Delta H(x,t) + \divergence \left(\nabla V(x) H(x,t)\right) + \sqrt{2D} \nabla \left(\sqrt{H(x,t)} Z^D(x,t) \right) \\ & \hspace{-1cm}+ \eta \sY(x,t) + \sqrt{\eta \sY(x,t)} Z_6^I(x,t)  - \phi_{h} H(x,t) 
   - \sqrt{\phi_{h} H(x,t)} Z_7^I(x,t) \\ & \hspace{-1cm}
   - \kappa H(x,t) - \sqrt{\kappa H(x,t)} Z_8^I(x,t) \\
   \frac{\partial C(x,t)}{\partial t} = \; &D \Delta C(x,t) + \divergence \left(\nabla V(x) C(x,t)\right) + \sqrt{2D} \nabla \left(\sqrt{C(x,t)} Z^D(x,t) \right) \\ & \hspace{-1cm}+ \kappa H(x,t) + \sqrt{\kappa H(x,t)} Z_8^I(x,t) - \eta_c C(x,t) - \sqrt{\eta_c C(x,t)} Z_9^I(x,t) \\
   \frac{\partial H_C(x,t)}{\partial t}=\; &D \Delta H_C(x,t) +\divergence \left(\nabla V(x) H_C(x,t)\right) + \sqrt{2D} \nabla \left(\sqrt{H_C(x,t)} Z^D(x,t) \right) \\ & \hspace{-1cm}+ \eta_c C(x,t) + \sqrt{\eta_c C(x,t)} Z_9^I(x,t) - \phi_{hc} H_C(x,t) 
    - \sqrt{\phi_{hc} H_C(x,t)} Z_{10}^I(x,t) \\
   \frac{\partial R(x,t)}{\partial t} = \; &D \Delta R(x,t) + \divergence \left(\nabla V(x) R(x,t)\right) + \sqrt{2D} \nabla \left(\sqrt{R(x,t)} Z^D(x,t) \right) \\ & \hspace{-1cm}+ \phi_i I(x,t) + \sqrt{\phi_i I(x,t)} Z_3^I(x,t) + \phi_{sy} \sY(x,t) + \sqrt{\phi_{sy} \sY(x,t)} Z_5^I(x,t) \\ & \hspace{-1cm}+ \phi_{h} H(x,t) 
     + \sqrt{\phi_{h} H(x,t)} Z_7^I(x,t) + \phi_{hc} H_C(x,t) + \sqrt{\phi_{hc} H_C(x,t)} Z_{10}^I(x,t),
\end{flalign*}
where $Z^D(x,t) \in \mathbb{R}^2$ denotes the white noise associated with diffusion and $Z_r^I(x,t) \in \mathbb{R}^2$ represents the white noise for the $r^{th}$ interaction, where $r=1,2,\dots,10$.
Since the expected value of the noise is zero, the resulting system of PDEs is given by 
\begin{equation}  \label{eq:PDE-model}
\begin{aligned} 
    \frac{\partial S(x,t)}{\partial t} = \; &D \Delta S(x,t) + \divergence (\nabla V(x) S(x,t)) \\ & - \beta S(x,t) \bigl( I(t)^{B_r(x)} + \sY(t)^{B_r(x)} \bigr) \\
    \frac{\partial E(x,t)}{\partial t} = \; &D \Delta E(x,t) + \divergence (\nabla V(x) E(x,t))  \\ & + \beta S(x,t) \bigl( I(t)^{B_r(x)} + \sY(t)^{B_r(x)} \bigr) - \sigma E(x,t)  \\
   \frac{\partial I(x,t)}{\partial t} = \; &D \Delta I(x,t) + \divergence (\nabla V(x) I(x,t)) + \sigma E(x,t) - \phi_i I(x,t) \\ &- \gamma I(x,t) \\
   \frac{\partial \sY(x,t)}{\partial t}= \; &D \Delta \sY(x,t)+ \divergence (\nabla V(x) \sY(x,t))+ \gamma I(x,t) \\ &- \phi_{sy} \sY(x,t) - \eta \sY(x,t) \\
   \frac{\partial H(x,t)}{\partial t} = \; &D \Delta H(x,t) + \divergence (\nabla V(x) H(x,t)) + \eta \sY(x,t)  - \phi_{h} H(x,t) \\ & - \kappa H(x,t) \\
   \frac{\partial C(x,t)}{\partial t} = \; &D \Delta C(x,t) + \divergence (\nabla V(x) C(x,t)) + \kappa H(x,t)                   - \eta_c C(x,t) \\
    \frac{\partial H_C(x,t)}{\partial t}=\; &D \Delta H_C(x,t)+\divergence (\nabla V(x) H_C(x,t))+ \eta_c C(x,t) \\ & - \phi_{hc} H_C(x,t)              \\
   \frac{\partial R(x,t)}{\partial t} = \; &D \Delta R(x,t) + \divergence (\nabla V(x) R(x,t)) + \phi_i I(x,t) + \phi_{sy} \sY(x,t) \\ & + \phi_{h} H(x,t) + \phi_{hc} H_C(x,t).
\end{aligned}
\end{equation}

Note that 
summing up all subpopulations results in a population movement corresponding to this PDE with a constant diffusion coefficient $D$
\begin{align*}
    \frac{\partial N(x,t)}{\partial t} = \; D \Delta N(x,t) + \divergence (\nabla V(x) N(x,t)),
\end{align*}
which is an approximation of our motion-based ABM 
for agents $i=1,2,\dots,N$ 
\begin{align*}
    dX_i(t) = -\nabla V(X_i(t)) dt + \sqrt{2D} \, dB_i(t). 
\end{align*}

\subsubsection{Weak Formulation} \label{sec:weakformulation}
To prepare the system of PDEs~\eqref{eq:PDE-model} for numerical simulation using the finite element method, we derive its weak formulation. This involves multiplying each equation by a test function $\varphi$ and integrating over the domain $\Omega_{Be}$. In that way, we get the weak formulation of the PDEs for the densities. For example, the PDE of susceptible density is then given by
\begin{align*}
    \int_{\Omega_{Be}} \frac{\partial S}{\partial t} \varphi \; dx 
    = \int_{\Omega_{Be}} & -D \nabla S \cdot \nabla \varphi 
    + \Delta V S \varphi 
    + \frac{\partial V}{\partial x_1} \frac{\partial S}{\partial x_1} \varphi 
    + \frac{\partial V}{\partial x_2} \frac{\partial S}{\partial x_2} \varphi \\ &
    - \beta S \biggl( \int_{B_r(x)} (I+\sY) \; dy \biggr) \varphi 
    \; dx.
\end{align*}
Assuming the Laplacian of our landscape vanishes almost everywhere, $\Delta V=0$ a.e., which follows from its histogram-based piecewise constant construction
, the equation reduces to
\begin{align*}
    \int_{\Omega_{Be}} \frac{\partial S}{\partial t} \varphi \; dx 
    = \int_{\Omega_{Be}} & -D \nabla S \cdot \nabla \varphi 
    + \frac{\partial V}{\partial x_1} \frac{\partial S}{\partial x_1} \varphi 
    + \frac{\partial V}{\partial x_2} \frac{\partial S}{\partial x_2} \varphi \\ & - \beta S \biggl( \int_{B_r(x)} (I+\sY) \; dy \biggr) \varphi 
    \; dx.
\end{align*}
Since our domain is quite large and the spatial mesh is relatively coarse, the distance between neighboring grid points is substantial -- on the order of hundreds of meters. In the context of infection spreading, this implies that within a ball $B_r(x)$ of contact radius $r$, it is reasonable to assume that only a single grid point contributes significantly. Hence, we will approximate the integral of infectious density over the ball $B_r(x)$ by the infectious density at the center $x$ scaled by the area of the ball 
\begin{align*}
\int_{B_r(x)} \bigl( I(y,t)+\sY(y,t) \bigr) \; dy \approx \pi r^2 \bigl( I(x,t)+\sY(x,t) \bigr).
\end{align*}
This localized approximation is consistent with the resolution of the spatial discretization and simplifies the infection term in the weak formulation. As such, the weak form for the susceptible compartment becomes:

\begin{align*}
    \int_{\Omega_{Be}} \frac{\partial S}{\partial t} \varphi \; dx 
    \approx  \int_{\Omega_{Be}}& -D \nabla S \cdot \nabla \varphi 
    + \frac{\partial V}{\partial x_1} \frac{\partial S}{\partial x_1} \varphi 
    + \frac{\partial V}{\partial x_2} \frac{\partial S}{\partial x_2} \varphi \\ & - \beta \pi r^2 S (I+\sY) \varphi 
    \; dx.
\end{align*}
Analogous approximations are applied to the weak formulations of the other compartments, which follow the same structure. For implementation details, we refer to the corresponding code (see \texttt{covid.hh}).
\section{Grid Creation} \label{appendix:grid}
We created a coarse grid of Berlin's modeling domain (see Fig.~\ref{fig:Berlin_grid_coarse}).
The mesh can be constructed via 
{\small{
\begin{lstlisting} 
triangle -pqa750000.0 <name of poly file without ending> 
\end{lstlisting}
}
for a mesh setup consisting of 
1,884 triangles,
1,053 nodes and
179 boundary nodes.
The -p switch loads a .poly file, which can define points, segments, holes, regional attributes, and area constraints (see \textit{Command line switches} in~\cite{triangle}).
With the -q switch, vertices are added to the mesh to ensure that no angle smaller than 20° occurs.
Using the -a switch, along with the value 750000.0, imposes a maximum triangle area of 750000.0.
Finally, the -p switch will generate a conforming constrained Delaunay triangulation, provided that either the -q or -a switch is also used.
\begin{figure}[h]
    \centering
    \includegraphics[width=0.3\linewidth]{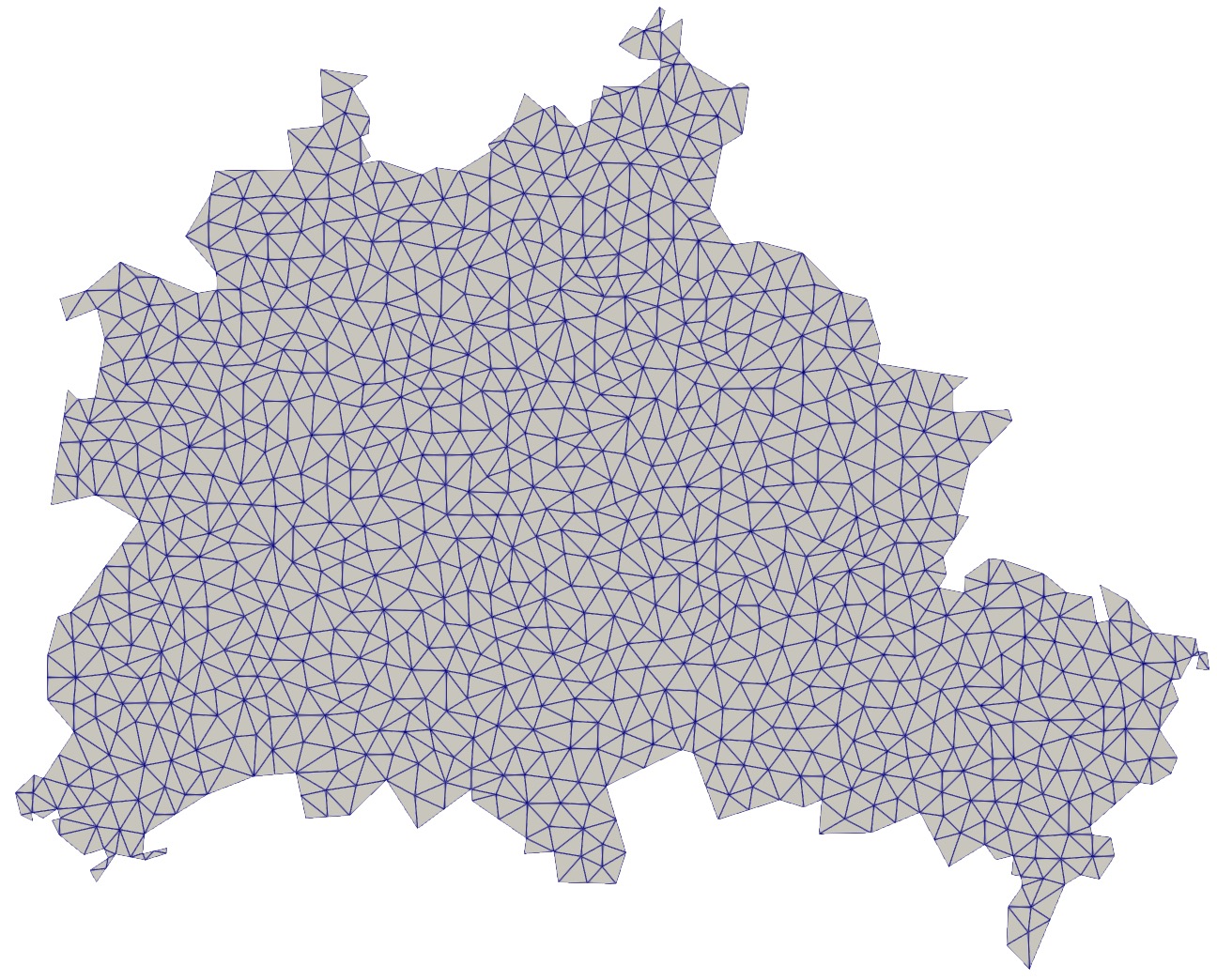}
    \caption{Modeling domain of PDE contribution (\ref{eq:PDE-model}) of hybrid model.}
    \label{fig:Berlin_grid_coarse}
\end{figure}

\section{Landscape Creation} 
\label{appendix:landscapeCreation}
%
Figures~\ref{fig:distribution_25pt} and~\ref{fig:distribution_100pt} illustrate the distribution of agents for the 25\% sample population and the full population, respectively. Figures~\ref{fig:landscape_25pt} and~\ref{fig:landscape_100pt} show the landscape of the agent distribution, including areas where no agent was ever present. In Figures~\ref{fig:landscape_berlin_before_transformation_25pt} and~\ref{fig:landscape_berlin_before_transformation_100pt}, these gaps were filled to prepare the data for the PDE model before transforming it from the rectangular grid to the triangular grid. Finally, Figures~\ref{fig:landscape_berlin_after_transformation_25pt} and~\ref{fig:landscape_berlin_after_transformation_100pt} display the landscape after the transformation.
\subsection{25\% Population}
\begin{figure}[h!]
    \centering
    \begin{minipage}{0.49\linewidth}
        \includegraphics[width=\linewidth]{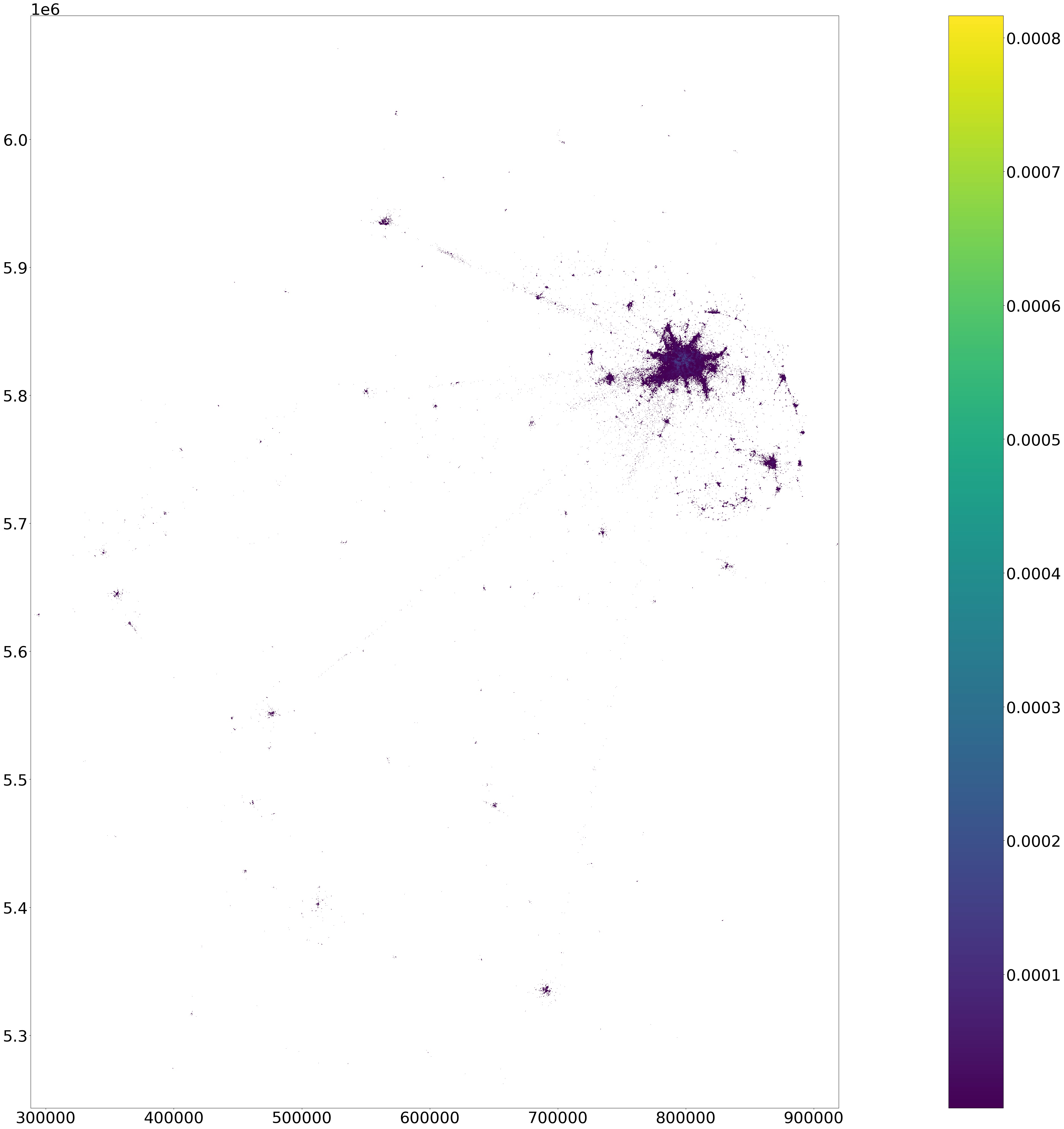}
    \end{minipage}
    \begin{minipage}{0.49\linewidth}
        \includegraphics[width=\linewidth]{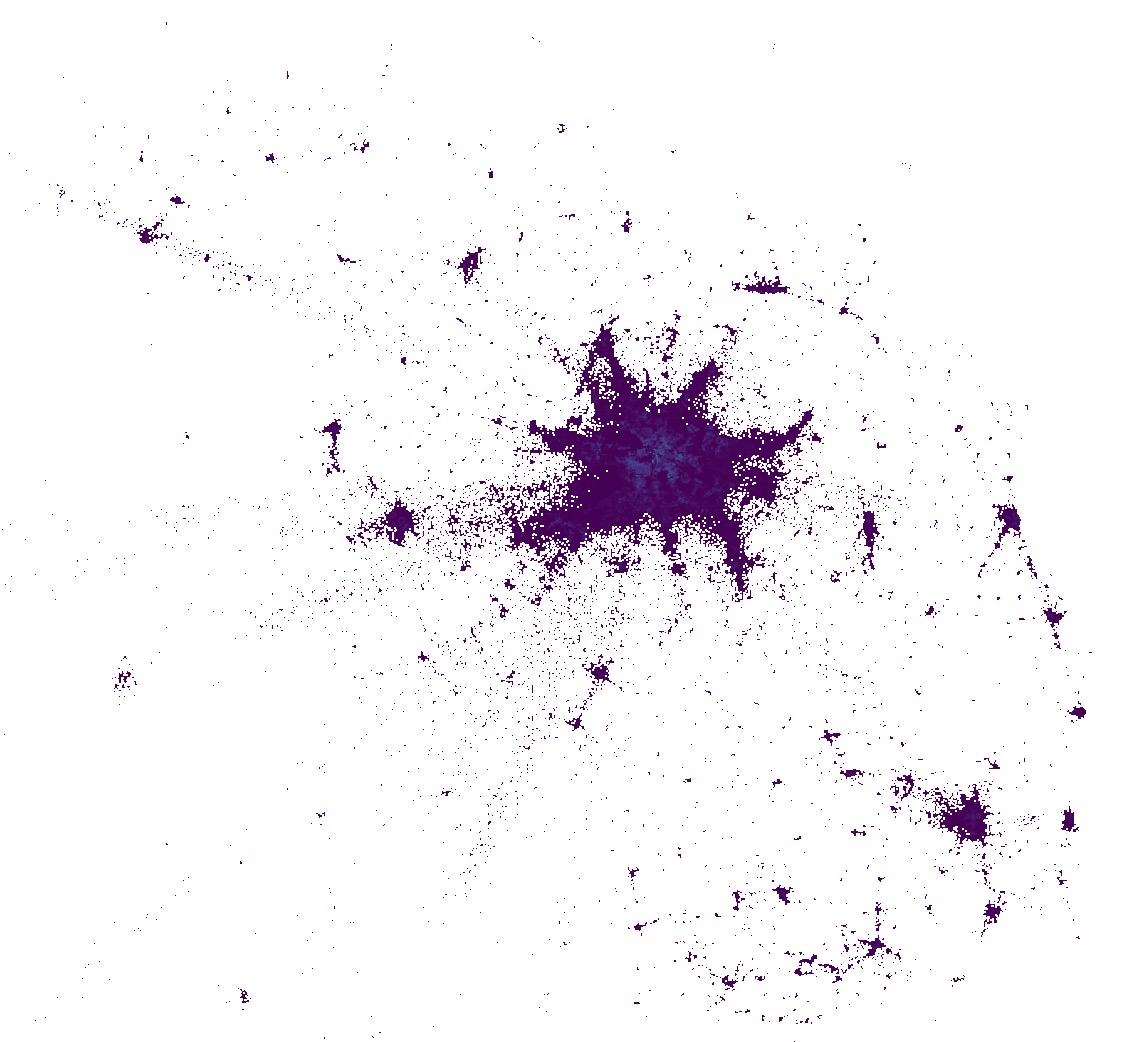}
    \end{minipage}
    \caption{Distribution of all agents (25\% population) across Germany for a whole week as a histogram, including NaN values. White color corresponds to NaN values, indicating that no agent has ever been at these coordinates. Right: Close up Berlin-Brandenburg. The close-up images were manually centered and therefore do not align perfectly.} 
    \label{fig:distribution_25pt}
\end{figure}
\begin{figure}[h!]
    \centering
    \begin{minipage}{0.49\linewidth}
        \includegraphics[width=\linewidth]{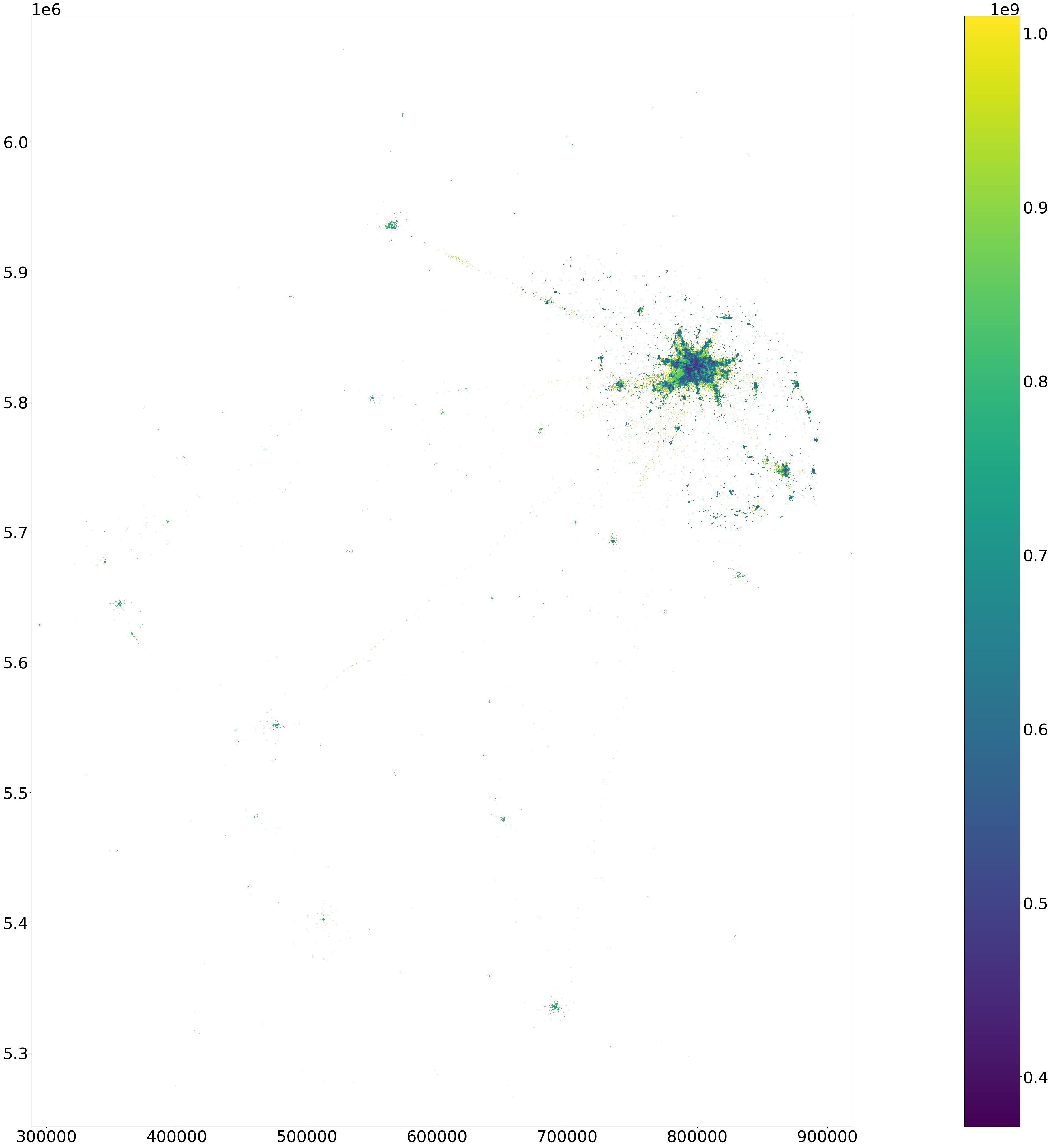}
    \end{minipage}
    \begin{minipage}{0.49\linewidth}
        \includegraphics[width=\linewidth]{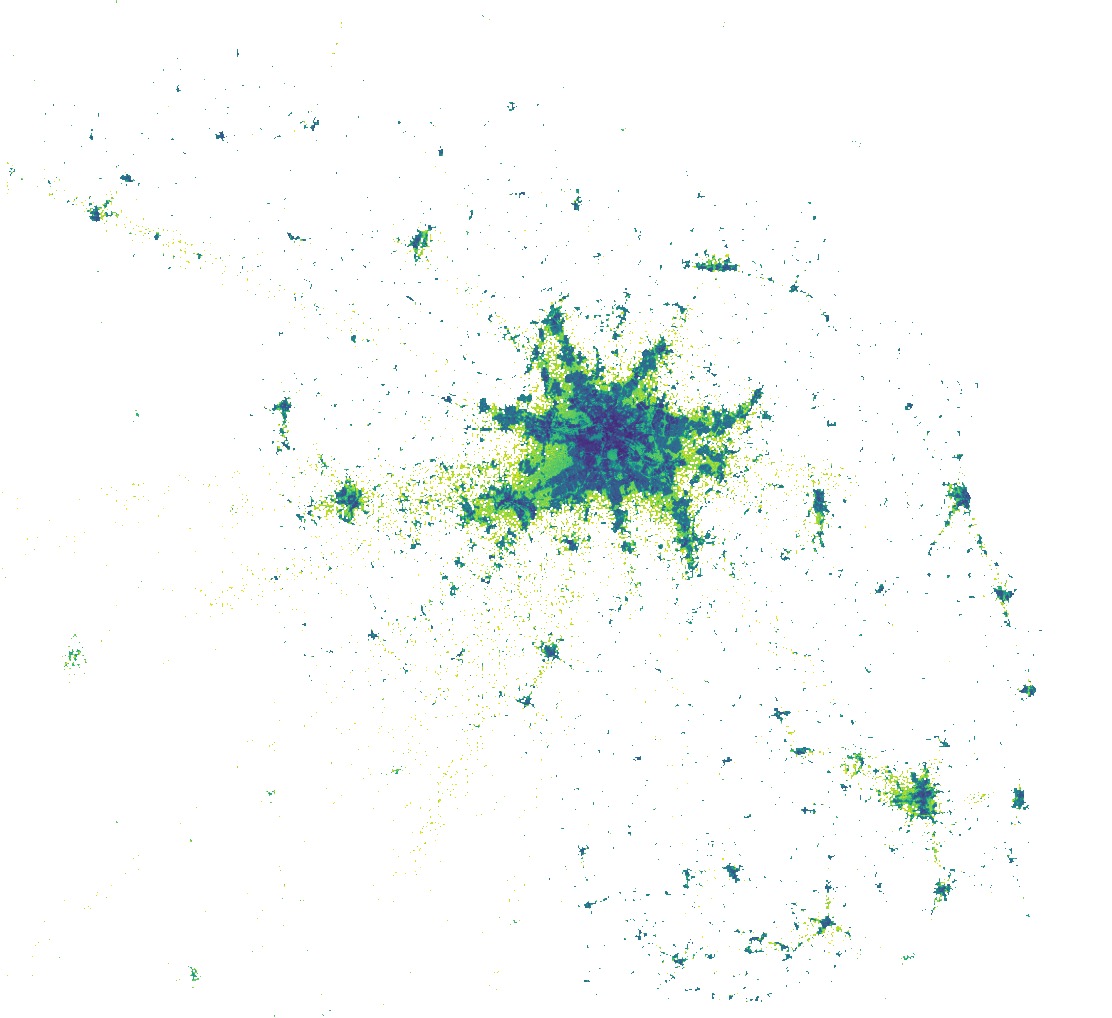}
    \end{minipage}
    \caption{Landscape of agent distribution (25\% population) across Germany over an entire week, including NaN values. White color corresponds to NaN values, indicating that no agent has ever been at these coordinates. Right: Close up Berlin-Brandenburg. The close-up images were manually centered and therefore do not align perfectly. }
    \label{fig:landscape_25pt}
\end{figure}
\begin{figure}[h!]
    \centering
    \begin{subfigure}{0.49\linewidth}
        \includegraphics[width=\linewidth]{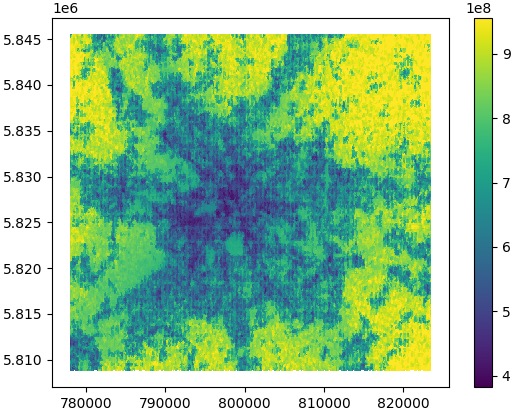}
        \caption{Landscape of rectangular area around Berlin on rectangular grid. 
        }
        \label{fig:landscape_berlin_before_transformation_25pt}
    \end{subfigure}
    \begin{subfigure}{0.49\linewidth}
        \includegraphics[width=\linewidth]{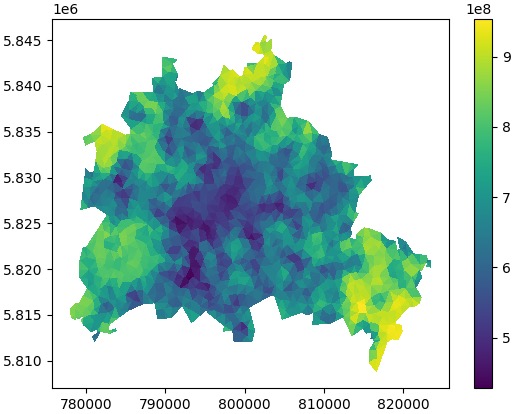}
        \caption{Landscape of Berlin on triangular grid. 
        }
        \label{fig:landscape_berlin_after_transformation_25pt}
    \end{subfigure}
    \caption{Landscape of agent distribution (25\% population) across Berlin over an entire week. Coordinates never visited by agents (NaN values) were set to the maximum value.}
\end{figure}
%
%
%
\newpage
\textcolor{white}{}
\newpage 
\textcolor{white}{}
\newpage 
\subsection{100\% Population}
\begin{figure}[h!]
    \centering
    \begin{minipage}{0.49\linewidth}
        \includegraphics[width=\linewidth]{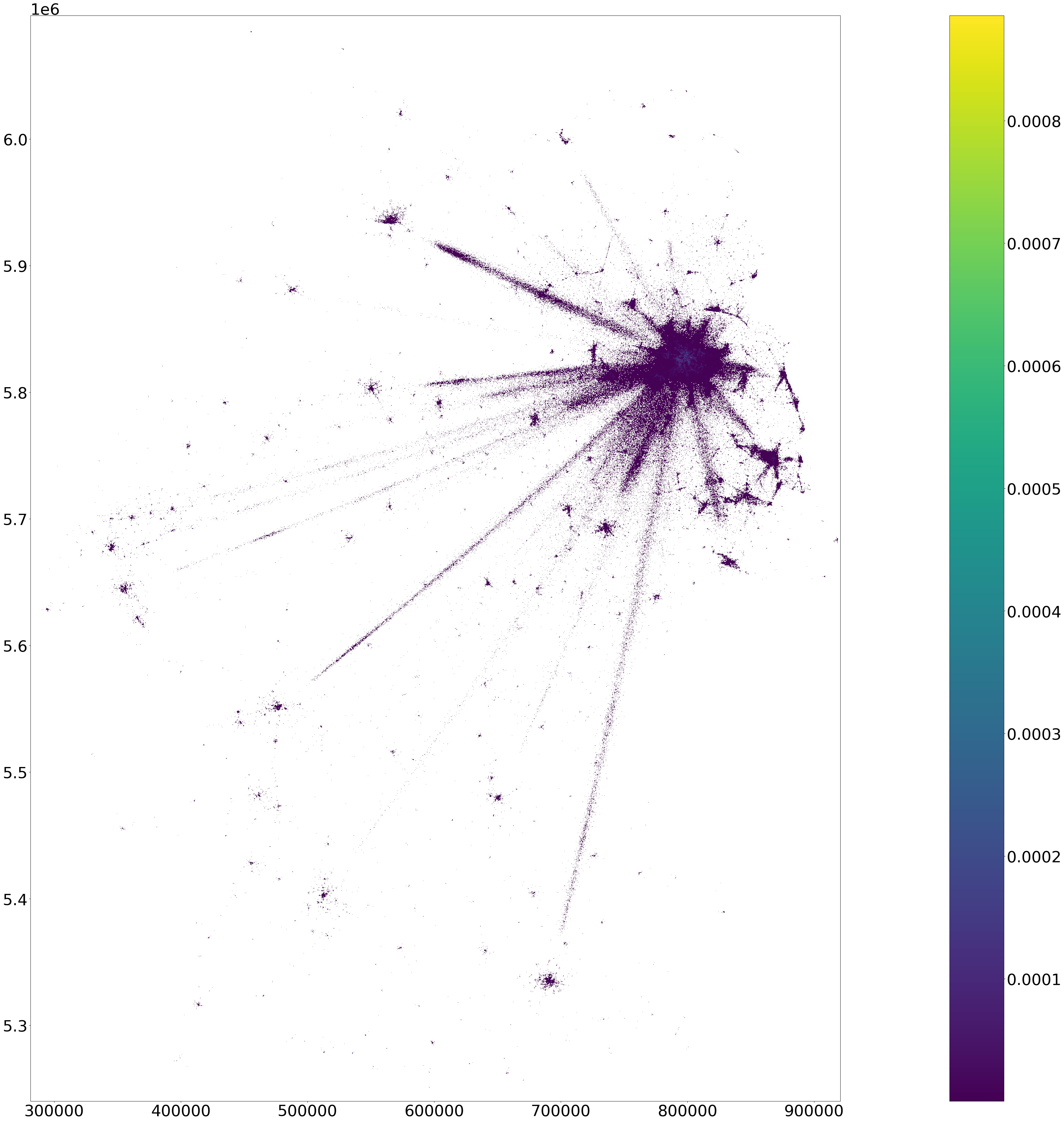}
    \end{minipage}
    \begin{minipage}{0.49\linewidth}
        \includegraphics[width=\linewidth]{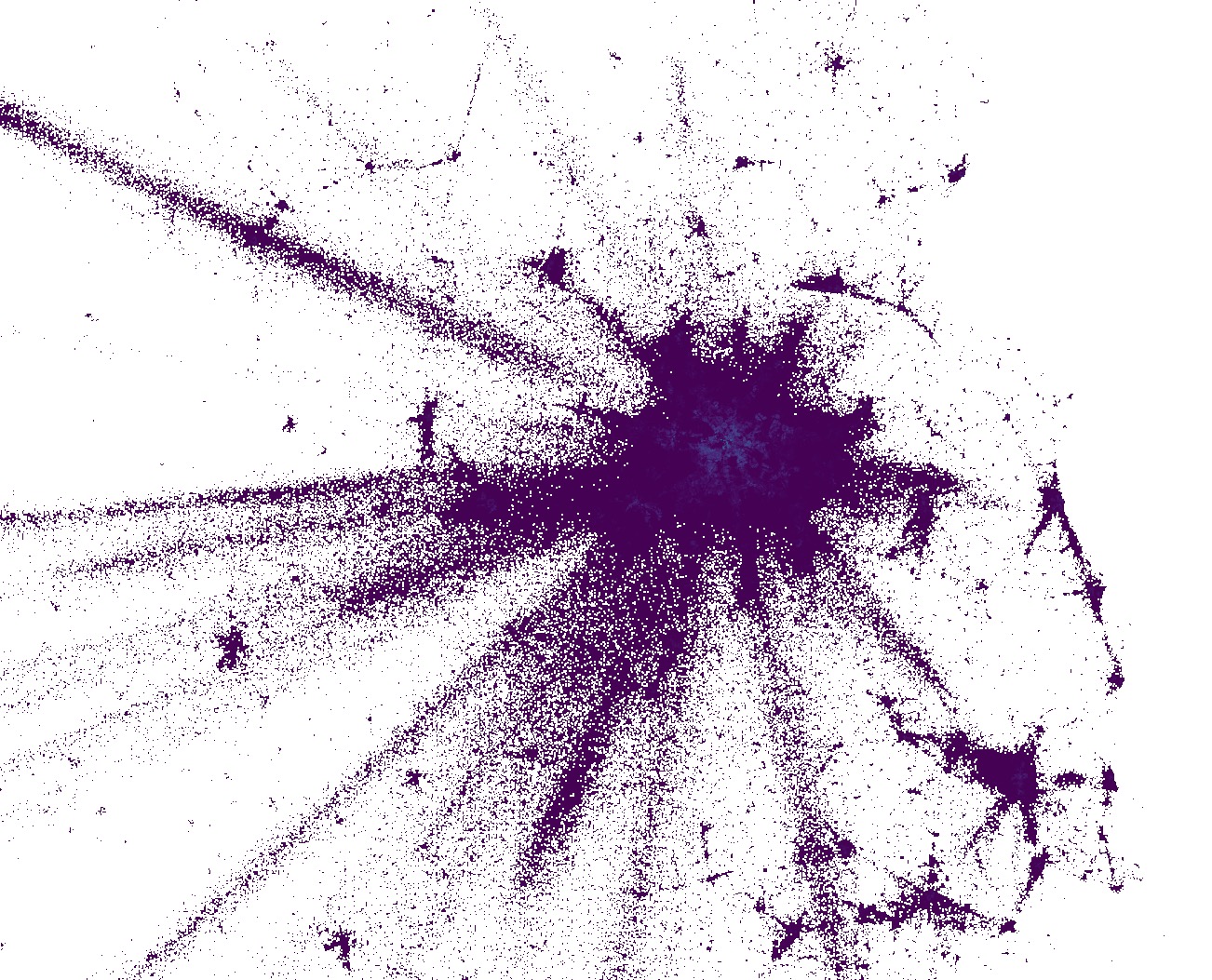}
    \end{minipage}
    \caption{Distribution of all agents (100\% population) across Germany for a whole week as a histogram, including NaN values. White color corresponds to NaN values, indicating that no agent has ever been at these coordinates. Right: Close up Berlin-Brandenburg. The close-up images were manually centered and therefore do not align perfectly.} 
    \label{fig:distribution_100pt}
\end{figure}
\begin{figure}[h!]
    \centering
    \begin{minipage}{0.49\linewidth}
        \includegraphics[width=\linewidth]{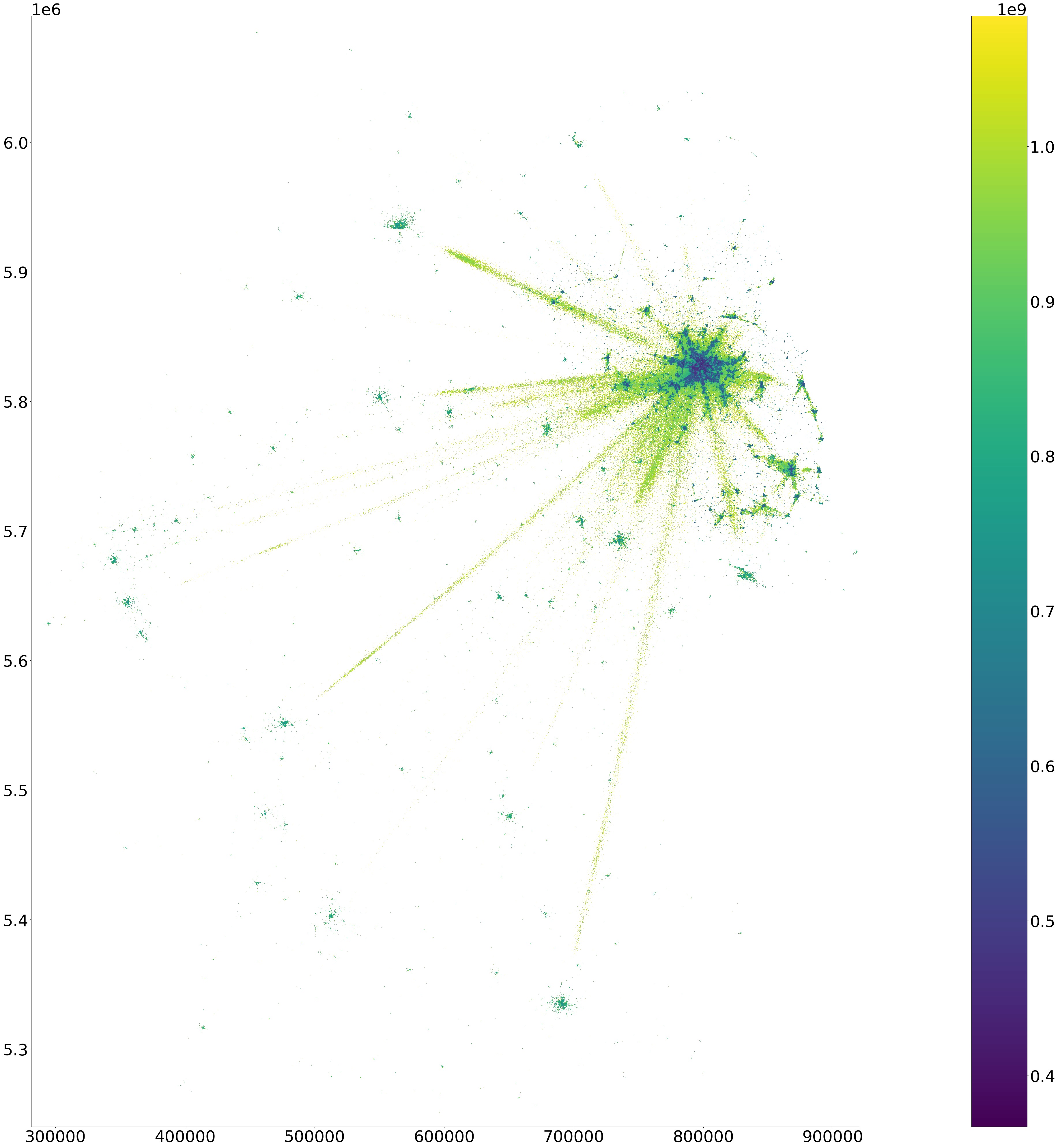}
    \end{minipage}
    \begin{minipage}{0.49\linewidth}
        \includegraphics[width=\linewidth]{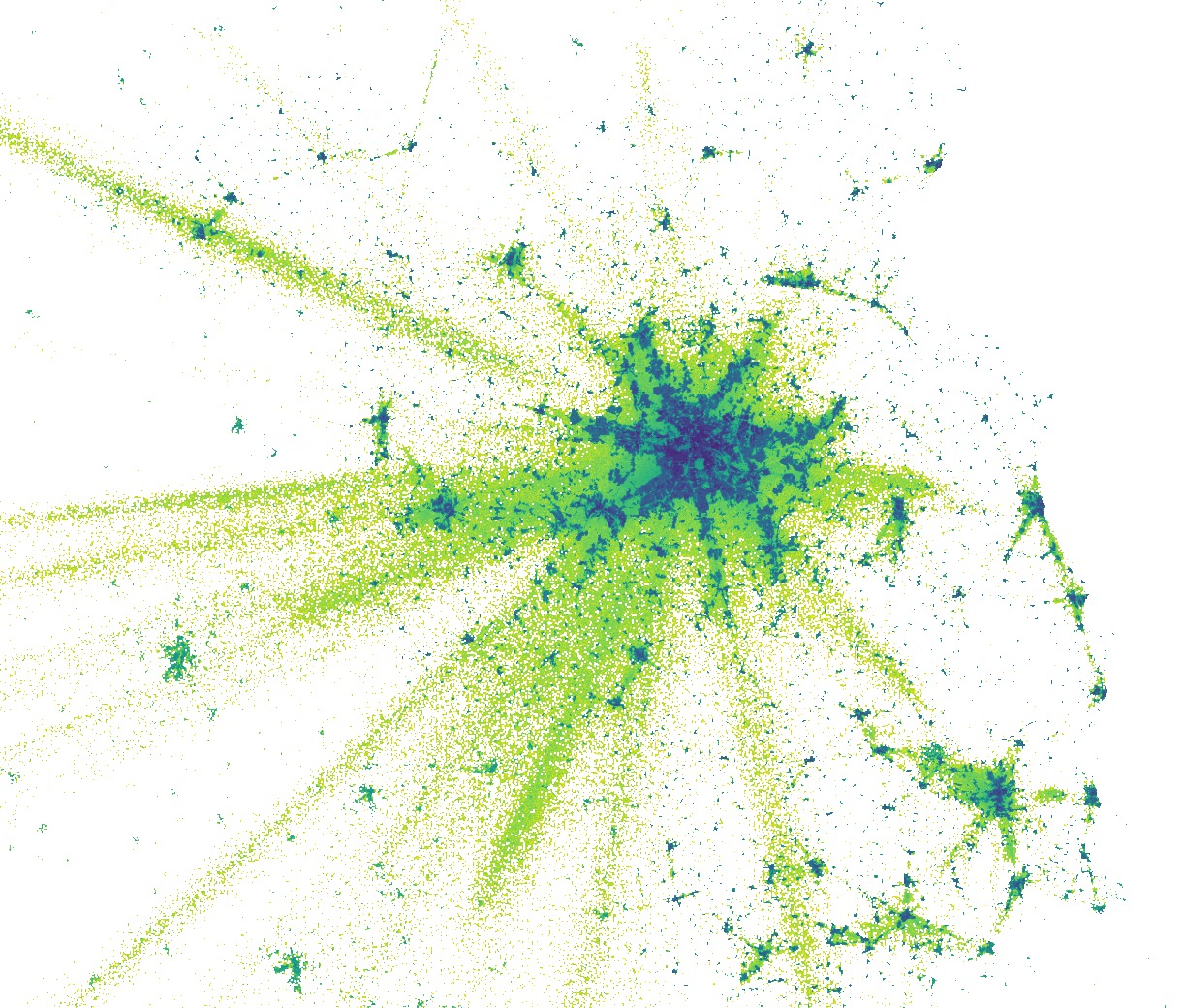}
    \end{minipage}
    \caption{Landscape of agent distribution (100\% population) across Germany over an entire week, including NaN values. White color corresponds to NaN values, indicating that no agent has ever been at these coordinates. Right: Close up Berlin-Brandenburg. The close-up images were manually centered and therefore do not align perfectly. }
    \label{fig:landscape_100pt}
\end{figure}
\begin{figure}[h!]
    \centering
    \begin{subfigure}{0.49\linewidth}
        \includegraphics[width=\linewidth]{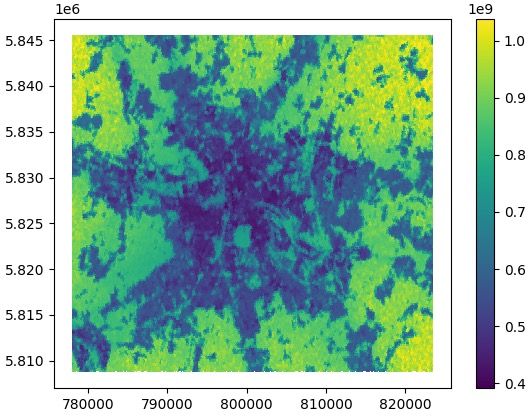}
        \caption{Landscape of rectangular area around Berlin on rectangular grid. 
        }
        \label{fig:landscape_berlin_before_transformation_100pt}
    \end{subfigure}
    \begin{subfigure}{0.49\linewidth}
        \includegraphics[width=\linewidth]{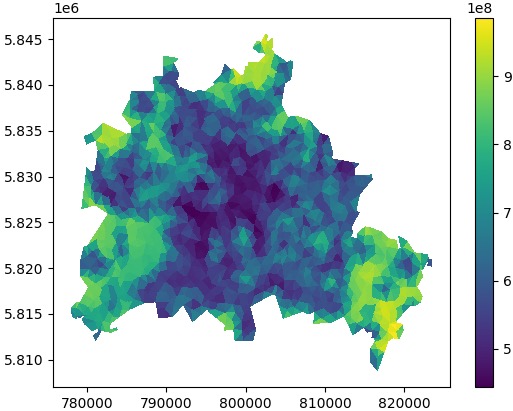}
        \caption{Landscape of Berlin on triangular grid. 
        }
        \label{fig:landscape_berlin_after_transformation_100pt}
    \end{subfigure}
    \caption{Landscape of agent distribution (100\% population) across Berlin over an entire week. Coordinates never visited by agents (NaN values) were set to the maximum value.}
\end{figure}

\newpage
\textcolor{white}{}
\newpage

\section{Number of Runs for Numerical Experiments} \label{appendix:NumberOfRuns}
We will examine the plots to observe when the cumulative runs stabilize. By applying specific thresholds for the relative changes between neighboring cumulative runs, we will analyze in detail how many runs are needed to reach the desired threshold for both the full-ABM and the hybrid model, for 25\% and 100\% of the population. This will be assessed in terms of both error and duration, with separate evaluations for each.

\subsection{25\% Population}
Figs.~\ref{fig:number_of_runs_plot_error_25pt_fullABM} and~\ref{fig:number_of_runs_plot_error_25pt_hybrid} show the cumulative absolute mean error of the ABM and hybrid model, respectively, based on 100 runs using a 25\% population sample.
\begin{figure}[h]
    \centering
    \begin{subfigure}{0.49\linewidth}
        \includegraphics[width=\linewidth]{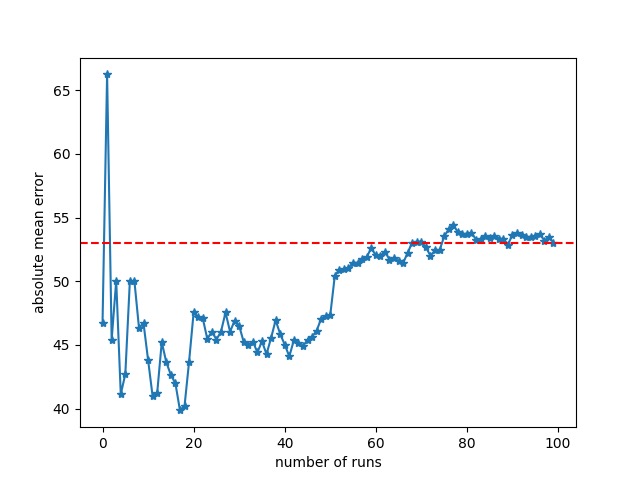}
        \caption{Full-ABM.} 
        \label{fig:number_of_runs_plot_error_25pt_fullABM}
    \end{subfigure}
    \begin{subfigure}{0.49\linewidth}
        \includegraphics[width=\linewidth]{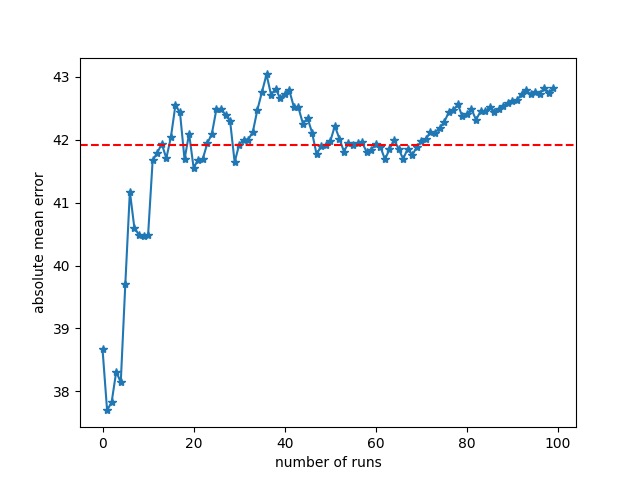}
        \caption{Hybrid model.} 
        \label{fig:number_of_runs_plot_error_25pt_hybrid}
    \end{subfigure}
    \caption{Cumulative absolute mean error of 100 runs using a 25\% population sample. Red line is mean of all cumulative runs.} 
\end{figure}
%
%
%
The relative difference between consecutive values is used as a threshold to determine the necessary number of runs.
With a 2\% relative difference between consecutive values, the hybrid model requires 13 runs, while the full-ABM requires 77 runs. Beyond 77 runs, the difference between consecutive values remains below 2\%, meaning additional runs (e.g., 78) do not significantly impact the result. This results in a 64-run reduction for the hybrid model.

For a 1\% relative difference, the hybrid model requires 31 runs, whereas the full-ABM needs 92 runs, leading to a 61-run reduction for the hybrid model.

Thus, halving the threshold roughly doubles the required runs for the hybrid model. Since the hybrid model already requires fewer runs than the full-ABM, this difference remains significant even for stricter thresholds, maintaining a substantial reduction in computational effort.

Computational time must be measured over multiple runs, as it can vary significantly depending on several factors. For example, the duration of individual disease courses can be highly variable. Additionally, reductions in activity participation can impact transmission dynamics by limiting interaction opportunities, potentially altering the computational load. By averaging across multiple runs, we ensure a more reliable estimate of computational time that accounts for these fluctuations. 

The cumulative mean running times of 10 runs are displayed in Figs.~\ref{fig:number_of_runs_plot_duration_25pt_fullABM} and~\ref{fig:number_of_runs_plot_duration_25pt_hybrid} for the full-ABM and the hybrid model, respectively. The hybrid model achieves a threshold of 0.56\% within the first two runs, whereas the ABM is initially close to 1.56\%. 
For a threshold of 0.2\%, the hybrid model achieves this again within 4 runs, while the ABM requires 5 runs.
The hybrid model reaches 0.01\% 
after just 9 runs, whereas the ABM's relative difference between the last two cumulative runs remains around 0.16\%.

\begin{figure}[h!]
    \centering
    \begin{subfigure}{0.49\linewidth}
        \includegraphics[width=\linewidth]{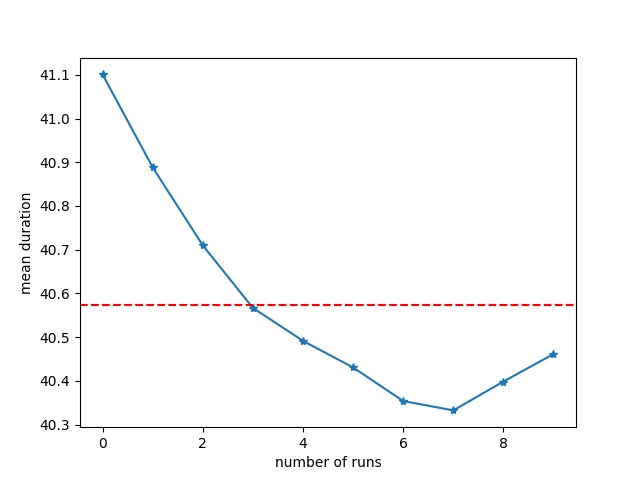}
        \caption{Full-ABM.} 
        \label{fig:number_of_runs_plot_duration_25pt_fullABM}
    \end{subfigure}
    \begin{subfigure}{0.49\linewidth}
        \includegraphics[width=\linewidth]{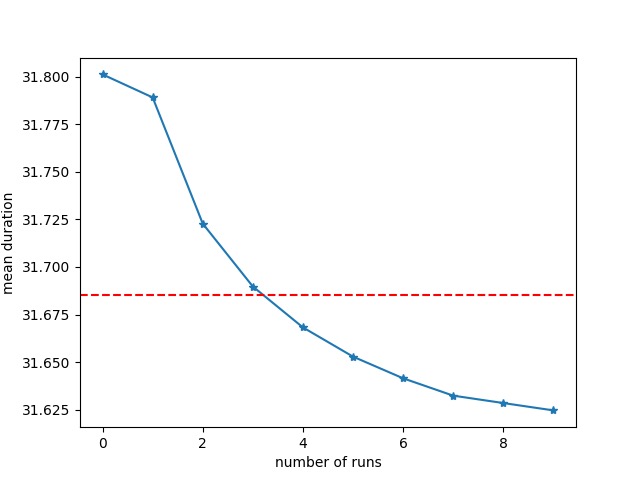}
        \caption{Hybrid model.} 
        \label{fig:number_of_runs_plot_duration_25pt_hybrid}
    \end{subfigure}
    \caption{Cumulative mean running times of 10 runs using a 25\% population sample. Red line is mean of all cumulative runs.} 
\end{figure}

\subsection{100\% Population}
The cumulative absolute mean error for the ABM and hybrid model, based on 100 runs with the full population sample (100\%), is shown in Figs.~\ref{fig:number_of_runs_plot_error_100pt_fullABM} and~\ref{fig:number_of_runs_plot_error_100pt_hybrid}, respectively. With a 2\% relative difference between consecutive values, the hybrid model requires only 4 runs, whereas the full-ABM needs 16. Beyond this point, additional runs (e.g., 17) have little effect, as the relative difference remains below 2\%, resulting in a 12-run reduction for the hybrid model.

For a 1\% relative difference, the hybrid model requires only 33 runs, while the full-ABM needs 44, leading to a 11-run reduction for the hybrid model. Notably, when the threshold is halved, the hybrid model requires about eight times as many runs, whereas the ABM’s run count almost triples. 
Again, the hybrid model shows an advantage over the ABM in terms of the required number of runs for these chosen thresholds. For smaller thresholds, however, this may not always be the case.

We observe a significant decrease in the number of runs required for the ABM when increasing the population from a 25\% sample to the full population. Assuming that the ABM for the full population takes four times longer to run than for the partial population, the choice of which sample to use depends on the selected threshold in combination with the number of runs and the hypothetical duration of a single run. For a threshold of 2\%, using the full population is clearly more efficient, whereas for a threshold of 1\%, the smaller sample proves to be the better option.

Interestingly, when examining the hybrid model results, the number of runs required for the full population either decreases or remains almost the same compared to the partial sample. In contrast, the ABM shows a noticeable decrease in the required number of runs. Regardless of the chosen threshold, this means that the hybrid model with the smaller population sample is, overall, probably the better choice for saving total simulation time to achieve stable results, assuming that the model for the full population will be slower than for the partial sample. For a proper efficiency comparison between the hybrid model and the ABM, the exact running times per single run are required.

\begin{figure}[h!]
    \centering
    \begin{subfigure}{0.49\linewidth}
        \includegraphics[width=\linewidth]{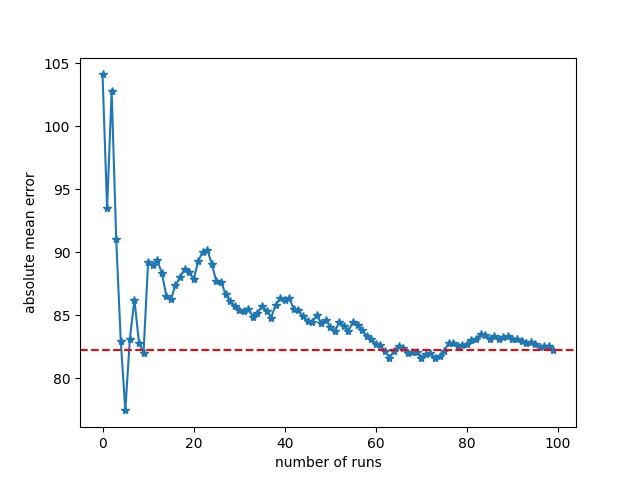}
        \caption{Full-ABM.} 
        \label{fig:number_of_runs_plot_error_100pt_fullABM}
    \end{subfigure}
    \begin{subfigure}{0.49\linewidth}
        \includegraphics[width=\linewidth]{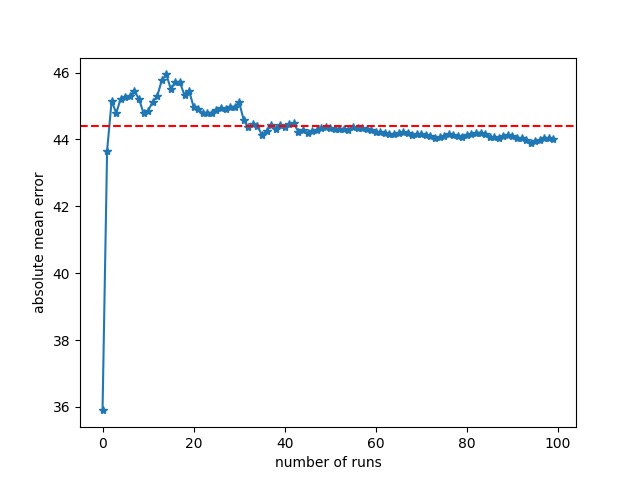}
        \caption{Hybrid model.} 
        \label{fig:number_of_runs_plot_error_100pt_hybrid}
    \end{subfigure}
    \caption{Cumulative absolute mean error of 100 runs using the full population sample (100\%). Red line is mean of all cumulative runs.} 
\end{figure}

Figs.~\ref{fig:number_of_runs_plot_duration_100pt_fullABM} and~\ref{fig:number_of_runs_plot_duration_100pt_hybrid} present the cumulative mean running times over 10 runs for the full-ABM and hybrid model, respectively. The hybrid model achieves a threshold of 0.08\% with only the first two runs. In contrast, the ABM is close to 3.09\%. 
For a 0.02\% relative difference, the hybrid model reaches the threshold with the first 6 runs, while the ABM requires more than 10 runs.

For a 0.01\% relative difference, the hybrid model achieves the desired threshold with just 10 runs. Specifically, the relative difference between the last two cumulative runs is 0.004\%. In contrast, the ABM requires more than 10 runs, with the relative change between the last consecutive runs being 0.21\%.

\begin{figure}[h!]
    \centering
    \begin{subfigure}{0.49\linewidth}
        \includegraphics[width=\linewidth]{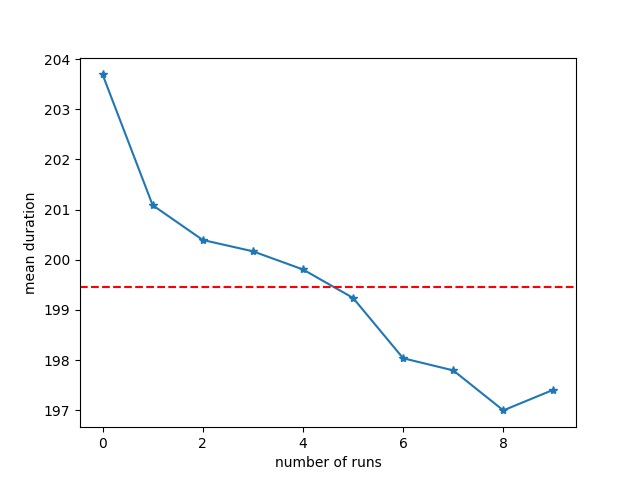}
        \caption{Full-ABM.} 
        \label{fig:number_of_runs_plot_duration_100pt_fullABM}
    \end{subfigure}
    \begin{subfigure}{0.49\linewidth}
        \includegraphics[width=\linewidth]{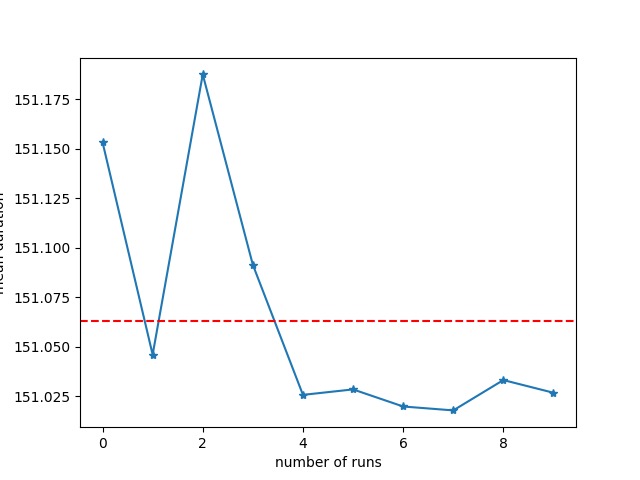}
        \caption{Hybrid model.} 
        \label{fig:number_of_runs_plot_duration_100pt_hybrid}
    \end{subfigure}
    \caption{Cumulative mean running times of 10 runs using the full population sample (100\%). Red line is mean of all cumulative runs.} 
\end{figure}
\end{document}